\documentclass[aps,prl,amsmath,amssymb, preprintnumbers, reprint,longbibliography,superscriptaddress]{revtex4-1}
\pdfoutput=1
\usepackage{lmodern, graphicx, multirow, xcolor, adjustbox}
\usepackage[colorlinks=true, citecolor=blue, urlcolor=blue, linkcolor=blue, breaklinks=true, pdfpagelabels=false]{hyperref}

\makeatletter\g@addto@macro\bfseries{\boldmath}\makeatother
\DeclareMathSymbol{\shortminus}{\mathbin}{AMSa}{"39}

\catcode`\$=\active
\gdef$#1${\texorpdfstring{\(#1\)}{\detokenize{#1}}}

\usepackage{wasysym}
\usepackage{lipsum}
\usepackage{multirow}
\usepackage{verbatim}
\usepackage{amsmath}

\usepackage{cancel}
\usepackage{hyperref}
\usepackage{makecell}

\begin{document}

\newcommand{\cpthree}{Centre for Cosmology, Particle Physics and Phenomenology (CP3), Universit\'{e} catholique de Louvain, 1348 Louvain-la-Neuve, Belgium}

\author{C\'eline Degrande}
\email{celine.degrande@uclouvain.be}
\affiliation{\cpthree}

\author{Matteo Maltoni}
\email{matteo.maltoni@uclouvain.be}
\affiliation{\cpthree}

\preprint{IRMP-CP3-24-09}

\title{EFT observable stability under NLO corrections through interference revival}

\begin{abstract}
We illustrate the importance of interference revival, when higher-order corrections are included, by presenting LO and NLO differential cross sections and {\it K}-factors for three processes that are sensitive to the dimension-6 SMEFT operator $O_W$: $Z$-plus-two-jets ({\it Zjj}) through VBF, leptonic diboson $WZ$ and $W \gamma$ production. We show how lifting the interference suppression at LO, through suitable variables and cuts, is necessary to get reliable predictions at NLO. We also present bounds on $C_W/\Lambda^2$ obtained from these observables
\end{abstract}

\maketitle

\subparagraph{Introduction}
Despite the achievements and success of accelerator physics in the past decades, no evidence for new resonances seems to be in sight in the near future; the presence of possible new heavy states can be investigated by searching for small anomalies in the interactions among the Standard Model (SM) particles. The Standard Model Effective Field Theory (SMEFT) provides a general tool to parametrise deviations from the SM, by adding to it complete sets of higher-dimensional operators $O_i$ with coefficients $C_i$ \cite{Grzadkowski:2010es,Buchmuller:1985jz}, 
\begin{equation}
\mathcal{L}_{\small{SMEFT}} = \mathcal{L}_{\small{SM}} + \sum_{i} \frac{C_i}{\Lambda^2}\hspace{1mm}O_i + \mathcal{O}(1/\Lambda^4),
\end{equation}
with $\Lambda$ a cutoff scale. The same expansion is observed in the differential cross section for a generic measurable variable $X$,
\begin{equation}
\frac{d\sigma}{dX} = \frac{d\sigma^{SM}}{dX} + \sum_i  \frac{C_i}{\Lambda^2}\frac{d\sigma^{1/\Lambda^2}}{dX}+ \mathcal{O}(1/\Lambda^4).
\end{equation}
It has been observed that the second term, being an interference between the SM amplitudes and the ones that are linear in $C_i$, can be suppressed for 2 $\rightarrow$ 2 processes \citep{Azatov:2016sqh} and in higher-multiplicity ones, resulting in the constraints coming from the term quadratic in $C_i$. Inspired by \cite{Dixon:1994}, in a previous work \cite{Degrande:2021rev} we suggested that suitable distributions can revive the sensitivity to the interference, providing limits on the coefficient which are dominated by the leading ($\mathcal{O}(1/\Lambda^2)$) term. 

In this article, we show how the application of the same procedure is important to get stable predictions at Next-to-Leading Order in Quantum Chromodynamics (NLO QCD) for these operators. In particular, we focus on the $O_W$ one, for which the large and negative {\it K}-factors computed at $\mathcal{O}(1/\Lambda^2)$ level highlight the presence of a suppression at Leading Order (LO), lifted at NLO \cite{Degrande:2021autom}. The (differential) $K$-factor is defined as the ratio of the (differential) cross sections at NLO and LO. Our analysis considers three processes that are sensitive to the effects of this operator, namely the fully leptonic electroweak (EW) $Z$-plus-two-jets production through Vector Boson Fusion (VBF), the fully leptonic $W^\pm Z$ diboson production, and the leptonic $W^\pm \gamma$; we show the predictions for some relevant distributions in the SM and at linear and quadratic orders.

The approach described in this paper is more general and can be applied to any phase space cancellation of the interference, even outside the SMEFT. For example, if at least two operators affect one process, only one linear combination of them will impact the total interference cross section, while all the others will be suppressed. Our approach is thus necessary to fully constrain those orthogonal directions of the parameter space, for which a cancellation occurs at linear level.

This method and the quantities it introduces can also be used together with machine-learning algorithms to develop suitable variables to restore the interference term or to compute asymmetries \cite{CPasymmML:2022higgs,CPasymmML:2022ew}.

\section{Framework} \label{sec:framework}
The operator that we consider in this work is a dimension-6 CP-even one and is defined as
\begin{equation}
O_W = \epsilon^{IJK} W^{I,\nu}_{\mu} W^{J,\rho}_{\nu} W^{K,\mu}_{\rho},
\end{equation}
where $W^{I,\nu}_\mu$ is the EW field strength. It can directly contribute to diboson processes and triple gauge couplings.

\subparagraph{Definitions}
In our previous work \cite{Degrande:2021rev}, we introduced the integrable cross section for the interference, namely
\begin{equation}
\sigma^{|\text{int}|} = \int d\Phi \hspace{1mm} \left|\frac{d\sigma^{1/\Lambda^2}}{d\Phi}\right| = \lim_{N\to\infty} \sum_{i=1}^N |w_i|,
\end{equation}
as the total effect of the interference over the entire phase space $\Phi$. It can be computed as the sum of the absolute values of the weights from a large enough sample of $N$ Monte Carlo unweighted events. Like the other quantities introduced in this section, it has to be obtained at parton level, as NLO and parton shower (PS) effects can generate negative weights that are impossible to distinguish from the interference ones.
A comparison between this and the actual cross section $\sigma^{1/\Lambda^2}$ gives an estimate of the suppression that affects the second one. This quantity, though, is not measurable, as it requires the knowledge of initial states, neutrino momenta and jet flavours and helicities. The measurable cross section is thus defined as
\begin{equation}
\sigma^{|\text{meas}|} = \int d\Phi_{\text{meas}} \hspace{1mm} \left|\sum_{\{\text{um}\}} \frac{d \sigma^{1/\Lambda^2}}{d\Phi}\right|,
\end{equation}
where the sum (integral) is performed over the set of discrete (continuous) unmeasurable quantities $\{\text{um}\}$. This observable is derived using only the information that is available in experiments, so it can be considered as an upper bound for any asymmetry, built on kinematic variables by summing the absolute values of the bin contents of their distributions, with the aim to restore the interference. It can be estimated over a sample of $N$ Monte Carlo events as
\begin{equation}
\sigma^{|\text{meas}|}=\lim_{N \rightarrow \infty} \sum_{i=1}^{N} \hspace{1mm} w_i \cdot \text{sign} \left( \sum_{\{\text{um}\}} \text{ME}(\vec{p}_i,\{\text{um}\}) \right),
\end{equation}
where ME is the interference part of the squared amplitude, $w_i$ is the weight of the $i^{th}$ event and $\vec{p}_i$ are the momenta of its final states. This quantity is computationally expensive to obtain, so our aim is to find easily-measurable kinematic variables that can approximate its value, in a general way that could be applied even outside the SMEFT framework or in regions where the EFT validity is questionable.

For each LO interference distributions, we also plot the relative cancellation
\begin{equation}
   R_{w\pm} = \frac{N_{w+}-N_{w-}}{N_{w+}+N_{w-}}, \label{Rwpm}
\end{equation}
defined as the difference between the numbers of positive and negative weights in each bin, divided by their sum; the cancellation is larger where this ratio is closer to zero.

\subparagraph{Generation details}
Our analysis is performed via {\sc MadGraph5}\_a{\sc MC@NLO} v3.4.2 \cite{Alwall:2014}, which we feed the SMEFT@NLO \cite{Degrande:2021autom} Universal FeynRules Output (UFO) \cite{Degrande:2011ua}, written from a FeynRules model \cite{Alloul:2013bka} that contains most of the CP-even dimension-6 operators in the SMEFT; {\sc NloCT} is used to get the rational and ultraviolet counterterms \cite{Degrande:2015nloct}. The leptons and quarks, except the top, are considered as massless. We use the NNPDF3.0 parton distribution function (PDF) set \cite{Ball:2015}, with $\alpha_s (M_Z)=0.118$. $C_W/\Lambda^2$ is fixed to 1 TeV${}^{-2}$ throughout the paper. We present predictions for the Large Hadron Collider (LHC) at 13 TeV.

For the calculations at NLO that are carried out at fixed order (FO), we set the renormalisation and factorisation scales to $\mu_R = \mu_F = 1$ TeV; the events generated at NLO \cite{Frixione:2002}, on the other hand, are showered through \textsc{Pythia8} \cite{Pythia:2015} or \textsc{Herwig7} \cite{Herwig:2008,Herwig:2016}. The sum of transverse energies divided by two ($H_T/2$) is chosen as dynamical scale. Dressed leptons are obtained through the $k_t$ algorithm with a radius parameter $R=0.1$ \cite{Fastjet:2012,kt_1,kt_2}, then jets are reconstructed from the remaining final states using the anti-$k_t$ algorithm with $R= 0.4$ \cite{AntiKt:2008}.

Numerical and scale uncertainties are reported for each result \cite{Frederix:2012}. Numerical errors are due to the limited number of events generated, while scale variations are computed by taking the envelope of nine scale combinations, in which $\mu_{R,F}$ are varied by factors 0.5 and 2.

\subparagraph{Bounds computation}
For given distributions, we derive limits on $C_W/\Lambda^2$ by considering the deviations of each theory, namely 
\begin{equation}
   x^{SM}_{\text{best}}+\frac{C_W}{\Lambda^2} x^{1/\Lambda^2} \left( + \frac{C_W^2}{\Lambda^4} x^{1/\Lambda^4} \right),
\end{equation}
with the $2^{\text{nd}}$ and $3^{\text{rd}}$ terms at LO or NLO, from real data when possible, or from the best SM prediction $x^{SM}_{\text{best}}$. The $\mathcal{O}(1/\Lambda^4)$ term receives contributions from the square of amplitudes with one insertion of $O_W$ and from the interference among the SM and dimension-8 operators. The latter are not considered in this study due to their large number \cite{dim8_basis}, but they can produce comparable effects to the former term \cite{Azatov:2016sqh}. The bounds at that level are thus not fully meaningful and are shown only as a comparison to the linear ones and to test the EFT validity.

\begin{figure}
  \caption{\small{FO LO and NLO differential cross-section distributions for the transverse mass of the $WZ$ system, over all the phase space (\textit{top}) and when cuts on $p_T^Z$ and $\phi_{WZ}$ are applied (\textit{centre} and \textit{bottom}). The black (orange, green) line represents the SM, divided by 50 (interference, quadratic). The \textit{K}-factors are also shown, together with the relative cancellation for LO interference. In the first plot, we also report N$^2$LO results for the SM at FO and the experimental measurements, divided by 50. The last bin contains the overflow}} \label{fig:mTwz}
  \includegraphics[width=0.49\textwidth]{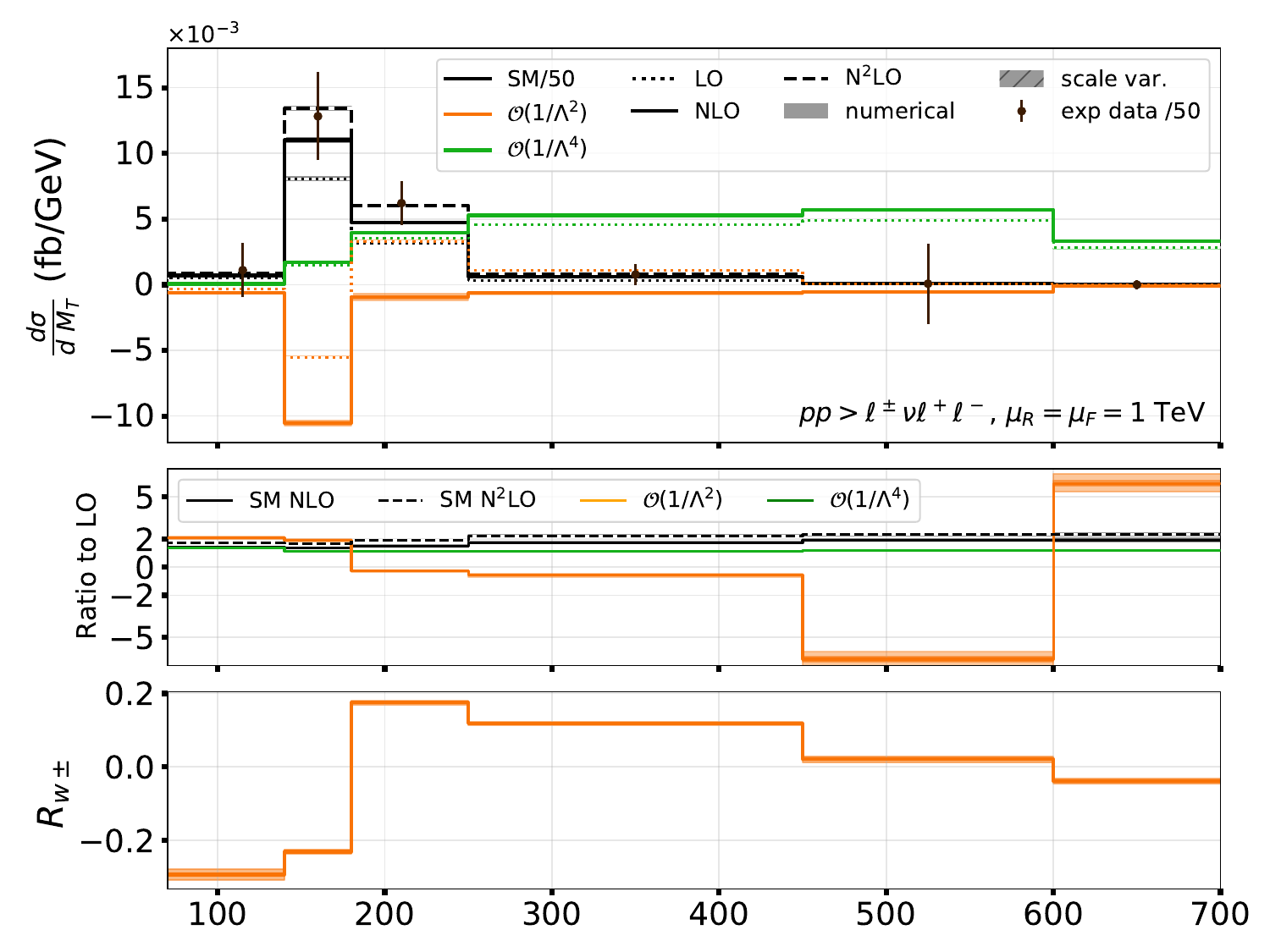}
  \includegraphics[width=0.49\textwidth]{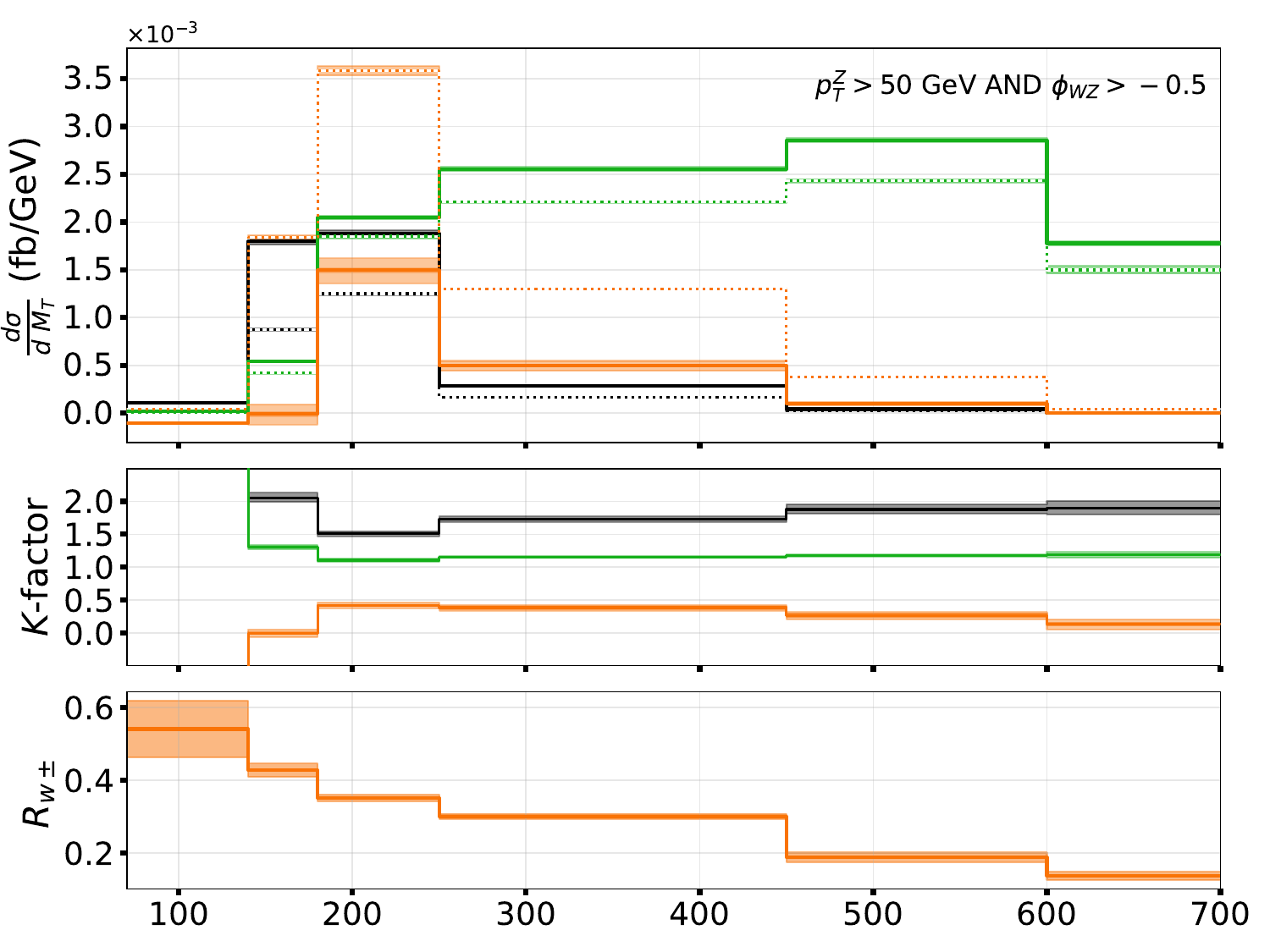}
  \includegraphics[width=0.49\textwidth]{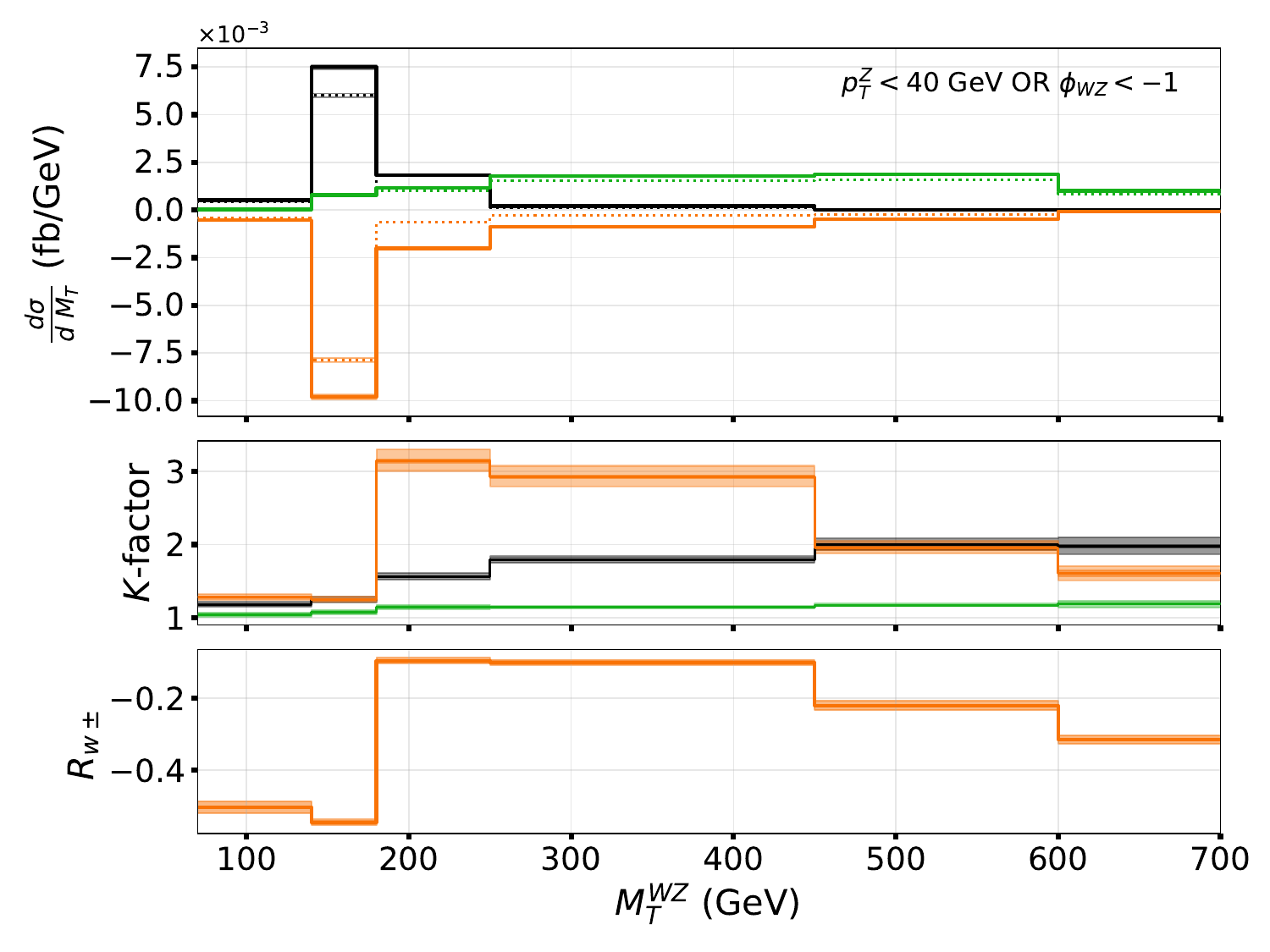}
\end{figure}

In these bound computations, we associate in each bin relative errors to the LO interference and $\mathcal{O}(1/\Lambda^4)$ contributions equal to $|k_i -1|$, with $k_i$ the $K$-factor for the bin; for negative values or above two, a 100\% uncertainty is considered, since the scale variations cannot be trusted as a good estimate of the missing higher-order corrections. For the SM and the terms at NLO, the numerical and scale uncertainties from our predictions are used, being the only available. As we will see for $WZ$ production in the top plot of Fig. \ref{fig:mTwz}, NLO uncertainties from {\sc MadGraph5} do not include the N${}^2$LO results inside them. This suggests that all the error bars shown in the plots in this paper, even the NLO ones for the SM, might underestimate the true uncertainties. In case of lack of real data, we assume that the experimental distribution follows the best SM one and we associate to it, in each bin $i$, a numerical uncertainty equal to $\sqrt{\sigma^{SM}_{\text{best},i}/\mathcal{L}_{LHC}}$, where the per-bin cross section is divided by the LHC luminosity at Run II, $\mathcal{L}_{LHC}= 137$ fb${}^{-1}$, plus a 10\% systematic one; no correlation among different bins is assumed.

\begin{figure}
   \caption{\small{NLO SM differential distributions for the azimuthal distance between jets in EW {\it Zjj}, at FO and matched with \textsc{Pythia8} and \textsc{Herwig7} showers. Numerical and scale-variation uncertainties are shown, and the experimental data as well}} \label{fig:dphijj_ps_comparison}
   \includegraphics[width=0.49\textwidth]{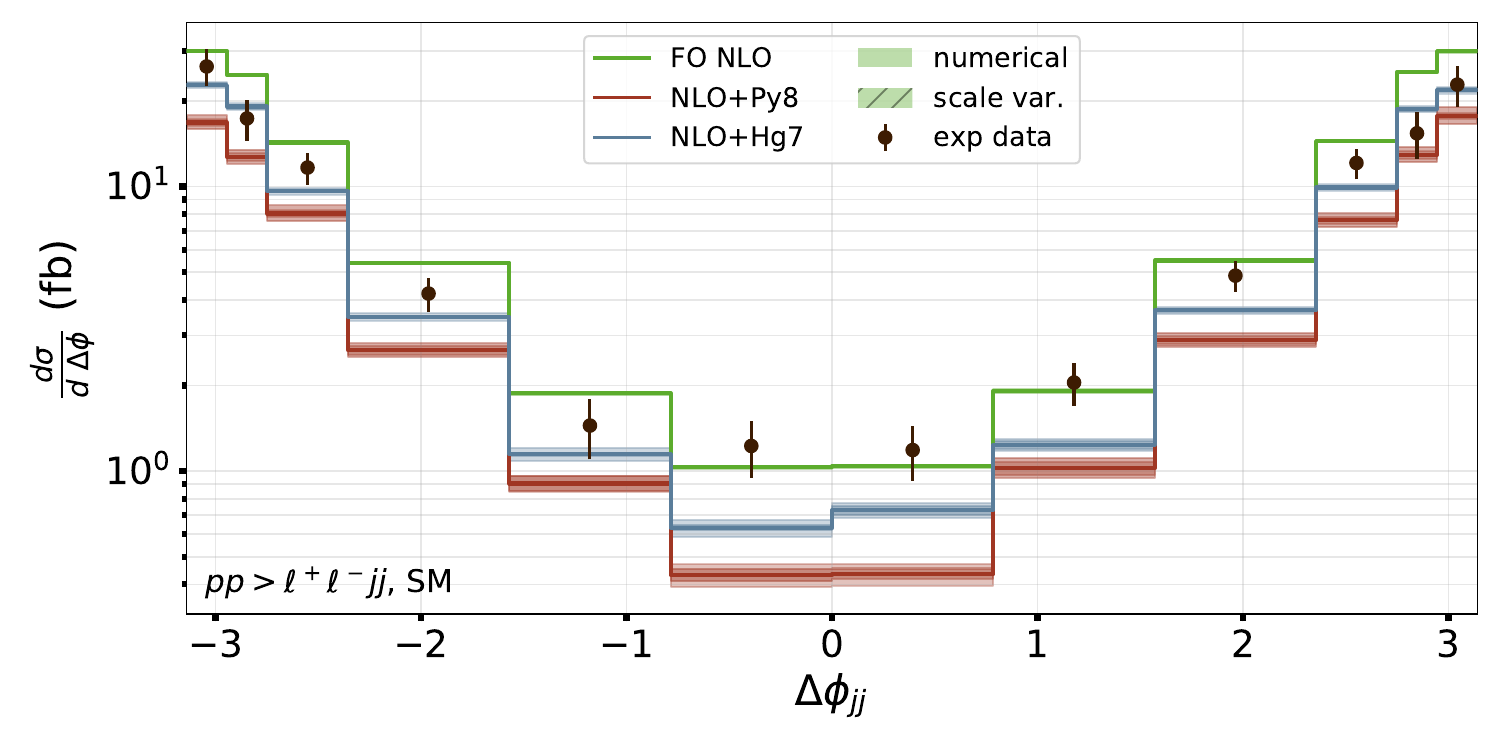}
\end{figure}

\begin{table*}
\caption{\small{Cross-section results in fb for the SM, linear and quadratic contributions, for the fully leptonic EW \textit{Zjj}, \textit{WZ} and $W \gamma$ production; the global \textit{K}-factors are also shown. $C_W/\Lambda^2$ is set to 1 TeV$^{-2}$. For each result, the first uncertainty source is numerical, while the second ones come from scale variation. For the \textit{K}-factors, the numerical uncertainty is propagated in quadrature from the cross-section ones, while for the scale variation the total envelope is considered. These results are obtained through FO computations for the first two processes, while the last one is matched to parton shower. The $WZ$ results are averaged over four decay channels}} \label{tab:xsect}
\begin{tabular}{c|ccc}
\hline
 & SM & $\mathcal{O}(1/\Lambda^2)$ & $\mathcal{O}(1/\Lambda^4)$ \\ \hline \hline
\multicolumn{4}{c}{$pp \rightarrow \ell^+ \ell^- jj$ EW, $\ell=(e,\mu)$} \\
\hline
$\sigma_{LO}$ (fb) & 49$\pm$0.06\%$^{+8\%}_{-6\%}$ & -1.67$\pm$0.4\%$^{+6\%}_{-7\%}$ & 9.4$\pm$0.07\%$^{+11\%}_{-10\%}$ \\
$\sigma_{NLO}$ (fb) & 52.2$\pm$0.19\%$^{+0.8\%}_{-1.1\%}$ & -1.66$\pm$1.2\%$^{+0.4\%}_{-0.8\%}$ & 11.1$\pm$0.18\%$^{+3\%}_{-4\%}$ \\
\textit{K}-factor & 1.07$\pm$0.19\%$^{+9\%}_{-7\%}$ & 0.99$\pm$1.2\%$^{+6\%}_{-8\%}$ & 1.18$\pm$0.17\%$^{+14\%}_{-14\%}$\\
\hline
\multicolumn{4}{c}{$pp \rightarrow \ell^\pm \overset{\scriptscriptstyle(-)}{\nu} \ell^+ \ell^-$, $\ell=(e,\mu)$} \\
\hline
$\sigma_{LO}$ (fb) & 34.6$\pm$0.012\%$^{+1.2\%}_{-1.4\%}$ & 0.169$\pm$0.3\%$^{+1.8\%}_{-2\%}$ & 6.2$\pm$0.06\%$^{+2\%}_{-1.6\%}$\\
$\sigma_{NLO}$ (fb) & 50.5$\pm$0.02\%$^{+1.6\%}_{-1.4\%}$ & -0.91$\pm$0.5\%$^{+5\%}_{-7\%}$ & 7.34$\pm$0.07\%$^{+0.8\%}_{-0.7\%}$\\
$\sigma_{N^2LO}$ (fb) & 62.8$\pm$0.3\%$^{+1.4\%}_{-1.3\%}$ & - & - \\
\textit{K}-factor & 1.46$\pm$0.03\%$^{+3\%}_{-3\%}$ & -5.4$\pm$0.6\%$^{+7\%}_{-9\%}$ & 1.18$\pm$0.09\%$^{+3\%}_{-3\%}$ \\
N$^2$LO / LO & 1.82$\pm$0.3\%$^{+3\%}_{-3\%}$ & - & - \\
\hline
\multicolumn{4}{c}{$pp \rightarrow \ell^\pm \overset{\scriptscriptstyle(-)}{\nu} \gamma$, $\ell=(e,\mu,\tau)$} \\
\hline
$\sigma_{LO}$ (fb) & 20.7$\pm$0.4\%$^{+1.4\%}_{-1.4\%}$ & -0.67$\pm$9\%$^{+21\%}_{-9\%}$ & 110$\pm$0.5\%$^{+5\%}_{-4\%}$ \\
$\sigma_{NLO}$ (fb) & 29.8$\pm$0.6\%$^{+3\%}_{-2\%}$ & -3.4$\pm$9\%$^{+9\%}_{-11\%}$ & 121$\pm$0.7\%$^{+1.2\%}_{-1.2\%}$ \\
\textit{K}-factor & 1.44$\pm$0.5\%$^{+4\%}_{-4\%}$ & 5.1$\pm$12\%$^{+29\%}_{-22\%}$ & 1.10$\pm$0.7\%$^{+6\%}_{-5\%}$ \\
\hline
\end{tabular}
\end{table*}

\begin{table}
\caption{\small{LO $\mathcal{O}(1/\Lambda^2)$ integral and measurable absolute-valued cross sections for {\it Zjj}, {\it WZ} and $W \gamma$, in fb. For the second process, the cases in which the $Z$-leptons helicities are separated are also shown, together with the regions in \eqref{cuts}. The asymmetries for some relevant variables are reported: they are the sum of the absolute values of the bin contents in their differential distributions. These results come from event generation at LO, without PS. The numerical uncertainties are shown: they are computed separately on the positive- and negative-weighted events, then propagated in quadrature}} \label{tab:meas}
\begin{tabular}{c|ccc}
\hline
 & (fb) & \% of $\sigma^{|\text{int}|}$ & \% of $\sigma^{|\text{meas}|}$ \\ \hline \hline
\multicolumn{4}{c}{$p p \rightarrow \ell^+ \ell^- j j$ EW, $\ell=(e,\mu)$} \\ \hline
$\sigma^{|\text{int}|}$ & 13.27$\pm$0.3\% & 100 & -\\
$\sigma^{|\text{meas}|}$ & 12.81$\pm$0.3\% & 97 & 100 \\
$\Delta \phi_{jj}$ & 11.42$\pm$0.4\% & 86 & 89 \\
$\sigma^{1/\Lambda^2}_{LO}$ & -1.71$\pm$2\% & 13 & 13 \\
\hline \hline
\multicolumn{4}{c}{$p p \rightarrow \ell^\pm \overset{\scriptscriptstyle(-)}{\nu} \ell^+ \ell^-$, $\ell=(e,\mu)$} \\ \hline
$\sigma^{|\text{int}|}$ & 4.93$\pm$0.4\% & 100 & - \\
$\sigma^{|\text{meas}|}$ & 2.04$\pm$1.0\% & 41 & 100\\
$p_T^Z \times \phi_{WZ}$ & 1.31$\pm$1.5\% & 27 & 64 \\
$\phi_{WZ}$ & 0.79$\pm$3\% & 16 & 39 \\
$M_T^{WZ}$ & 0.66$\pm$3\% & 13 & 32 \\
$\cos \theta^*_{\ell_Z^- Z}$ & 0.20$\pm$10\% & 4 & 10 \\
$\sigma^{1/\Lambda^2}_{LO}$ & 0.20$\pm$10\% & 4 & 10 \\
\hline
\multicolumn{4}{c}{$h (\ell_Z^-)=-1, h (\ell_Z^+) =+1$} \\ \hline
$\sigma^{|\text{int}|}$ & 2.773$\pm$0.5\% & 100 & - \\
$\sigma^{|\text{meas}|}$ & 1.738$\pm$0.9\% & 63 & 100 \\
$M_T^{WZ}$ & 0.38$\pm$4\% & 14 & 21 \\
$\sigma^{1/\Lambda^2}_{LO}$ & 0.108$\pm$14\% & 4 & 6 \\
\hline
\multicolumn{4}{c}{$h (\ell_Z^-)=+1, h (\ell_Z^+) =-1$} \\ \hline
$\sigma^{|\text{int}|}$ & 2.135$\pm$0.6\% & 100 & - \\
$\sigma^{|\text{meas}|}$ & 1.067$\pm$1.1\% & 50 & 100 \\
$M_T^{WZ}$ & 0.289$\pm$4\% & 14 & 27 \\
$\sigma^{1/\Lambda^2}_{LO}$ & 0.087$\pm$14\% & 4 & 8 \\
\hline
\multicolumn{4}{c}{$p_T^Z> 50$ GeV AND $\phi_{WZ}> -0.5$} \\
\hline
$\sigma^{|\text{int}|}$ & 2.260$\pm$0.7\% & 100 & - \\
$\sigma^{|\text{meas}|}$ & 0.873$\pm$1.7\% & 39 & 100 \\
$M_T^{WZ}$ & 0.660$\pm$2\% & 29 & 76 \\
$\sigma^{1/\Lambda^2}_{LO}$ & 0.660$\pm$2\% & 29 & 76 \\
\hline
\multicolumn{4}{c}{$p_T^Z< 40$ GeV OR $\phi_{WZ}< -1$} \\
\hline
$\sigma^{|\text{int}|}$ & 1.810$\pm$0.5\% & 100 & - \\
$\sigma^{|\text{meas}|}$ & 0.870$\pm$1.1\% & 48 & 100 \\
$M_T^{WZ}$ & 0.480$\pm$2\% & 27 & 55 \\
$\sigma^{1/\Lambda^2}_{LO}$ & -0.480$\pm$2\% & 27 & 55 \\
\hline
\hline
\multicolumn{4}{c}{$p p \rightarrow \ell^\pm \overset{\scriptscriptstyle(-)}{\nu} \gamma$, $\ell=(e,\mu,\tau)$} \\
\hline
$\sigma^{|\text{int}|}$ & 31.44$\pm$0.3\% & 100 & - \\
$\sigma^{|\text{meas}|}$ & 12.50$\pm$0.9\% & 40 & 100 \\
$\phi_W$ & 9.90$\pm$1.1\% & 31 & 79 \\
$p_T^\gamma \times |\phi_W|$ & 9.90$\pm$1.1\% & 31 & 79 \\
$p_T^\gamma \times |\phi_f|$ & 1.44$\pm$7\% & 5 & 12 \\
$\sigma^{1/\Lambda^2}_{LO}$ & -1.44$\pm$7\% & 5 & 12 \\
\hline
\end{tabular}
\end{table}

\begin{figure}
   \caption{\small{Differential cross section for the signed azimuthal distance between jets in {\it Zjj}, at LO without PS. The black line reproduces the SM distribution divided by 10, while the red (blue) one the positive- (negative-) weigthed contribution to the linear term. The orange line is the difference of the last two, namely the interference differential cross section. The uncertainties are not shown}} \label{fig:dphijj_lo}
   \includegraphics[width=0.49\textwidth]{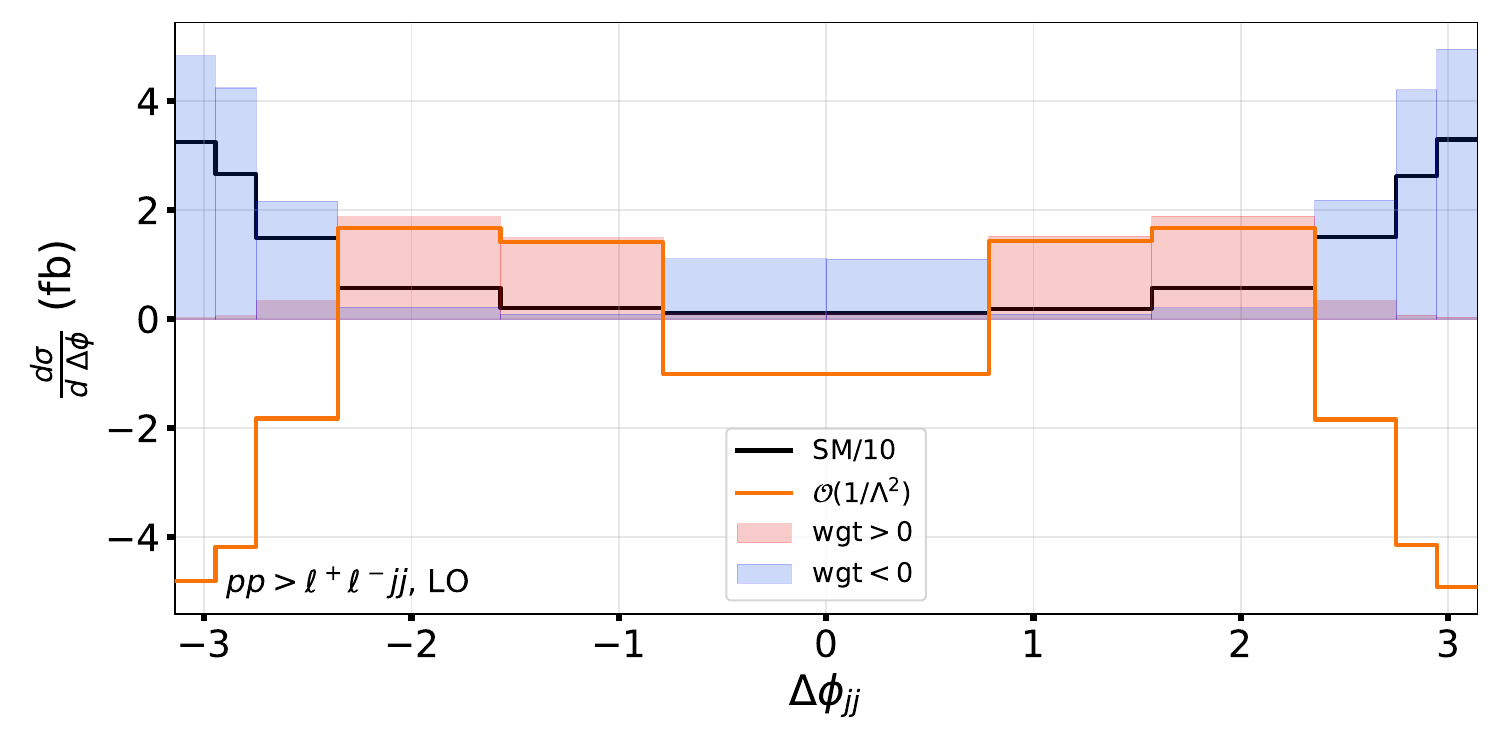}
\caption{\small{FO NLO ({\it continuous}) and LO ({\it dotted}) differential cross sections for $\Delta \phi_{jj}$ in {\it Zjj}, with the {\it K}-factors and the cancellation level \eqref{Rwpm} for the LO interference in each bin. The black (orange, green) line refers to the SM, divided by 10 (linear, quadratic terms); the numerical uncertainties and scale variations are also shown in the first two panels, while the last one only contains the numerical ones. The experimental measurements are represented by the dots with error bars, divided by 10}} \label{fig:dphijj_nlo}
   \includegraphics[width=.49\textwidth]{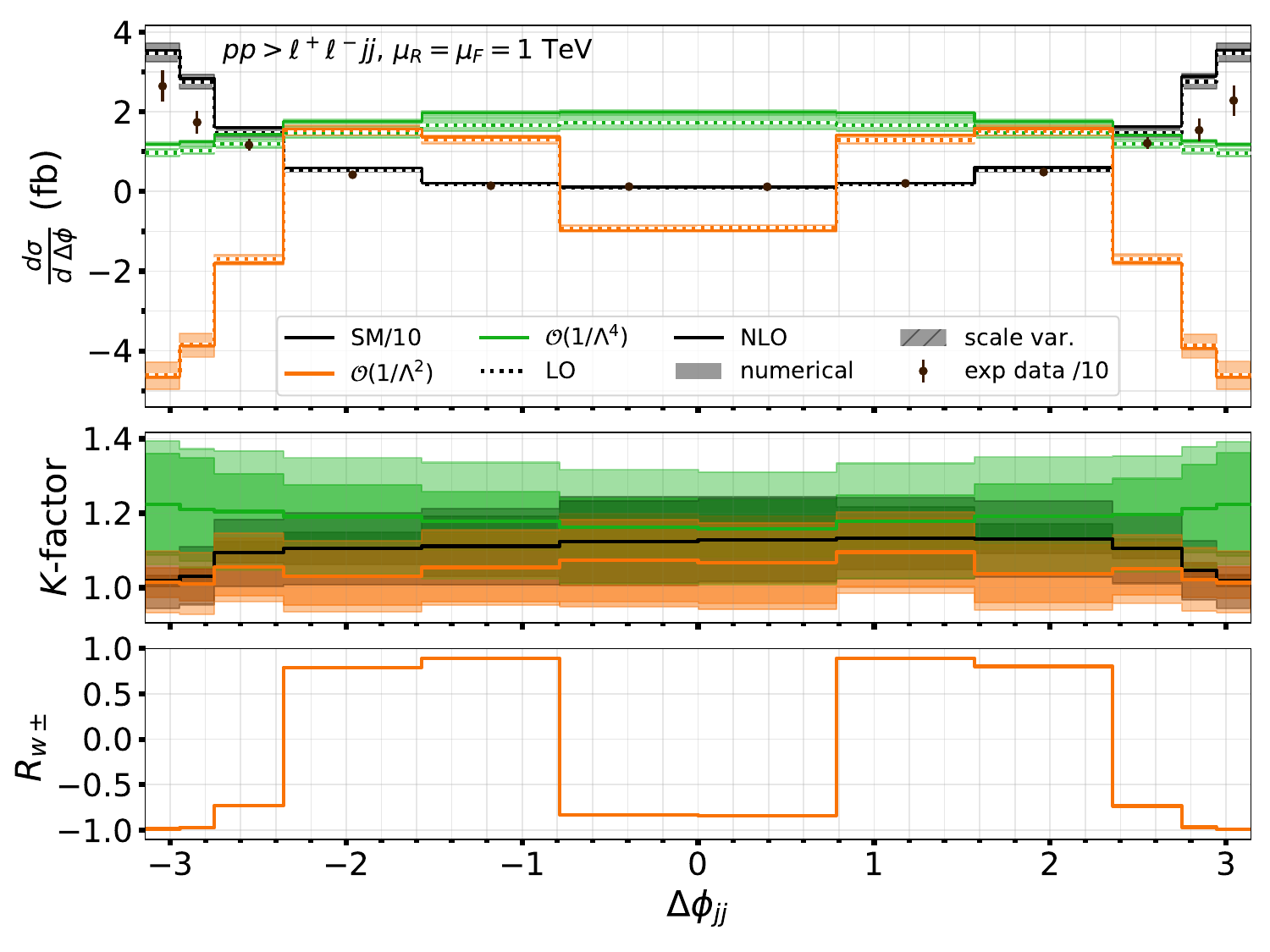}
\end{figure}

\section{Electroweak \textit{Z} + two jets production through Vector Boson Fusion}

\subparagraph{Calculational details}
The EW {\it Zjj} production is characterised by the exchange of a weak vector boson in the $t$-channel. We consider the $Z \rightarrow \ell^+ \ell^-$ decay case, with $\ell = e, \mu$. Following ATLAS specifications \cite{Atlas:2020zjj}, leptons are required to have transverse momentum $p_T > 25$ GeV and pseudorapidity $|\eta| < 2.4$; the total invariant mass of the leptons coming from the decay of the $Z$ boson has to satisfy $81.2 < M_{\ell \ell} < 101.2$ GeV, and their total transverse momentum $p_T^{\ell \ell} > 20$ GeV.

We require a leading jet with $p_T > 85$ GeV and a subleading one with $p_T> 80$ GeV. The rapidity for each jet has to be $|y|< 4.4$ and a separation is required among them and the leptons, namely $\Delta R_{\ell j}> 0.4$ (we consider $\Delta R = \sqrt{\Delta \phi^2 + \Delta \eta^2}$ with $\Delta \phi,\Delta\eta$ the azimuthal and pseudorapidity distances).

The EW contribution to the process is characterised by a large invariant mass for the jets, $M_{jj} > 1$ TeV, and a large gap in rapidity among them, $|\Delta y_{jj}| > 2$. No other jets with $p_T > 25$ GeV can be found in this rapidity interval. Furthermore, we impose to the $Z$ boson to be centrally produced relatively to the dijet system, by asking $\xi_Z < 0.5$; the last quantity is defined as
\begin{equation}
   \xi_Z = \frac{|y_{\ell \ell} - \frac{1}{2}(y_{j1}-y_{j2})|}{|\Delta y_{jj}|},
\end{equation}
with $y_{\ell \ell}$, $y_{j1}$ and $y_{j2}$ the rapidities of the dilepton system, the leading and subleading jets.

Different orders in the fine-structure constants $\alpha_S$ and $\alpha_W$ can be identified at LO and NLO for this process, and some of the NLO ones can both be seen as QED and QCD corrections to different LO ones. In order to ensure the cancellation of poles without computing all the orders, as it would require too much computing power, we asked for $W$ bosons only to be exchanged along the $t$-channel. In this way, the two external quark lines in the diagrams cannot feature the same flavour, making the NLO correction in QCD to the pure-EW component well-defined without having to include the LO interference among it and the QCD contribution. To implement this, we modified the model to include a coupling for the $W q \bar q$ vertex, and used it to reject $Z$ bosons and photons in the $t$-channel. We checked that the total LO SM cross section changes by less than 1\%, for the cuts above.

\subparagraph{Results}
The total LO and NLO cross sections at FO for the SM, linear and quadratic contributions can be found in Table \ref{tab:xsect}, together with the relative $K$-factors. Our FO SM NLO fiducial total cross section overestimates the ATLAS one by a factor $\sim 1.5$: it is known that predictions from different generators do not agree for VBF processes \cite{Atlas:2020zjj, ATL-PHYS-PUB-2021-022,WWjj_2019,WWjj_1_2019} and that results heavily depend on the shower choice \cite{Hoche:2021,Jager:2020}.

An analysis by the ATLAS Collaboration \citep{Atlas:2020zjj,ATL-PHYS-PUB-2021-022} showed that the signed-azimuthal angle difference between the two jets is particularly sensitive to the $O_W$ effects. This variable is defined as $\Delta \phi_{jj} = \phi_{ja} - \phi_{jb}$, where the jets are ordered in rapidity so that $y_{ja} > y_{jb}$.

In Fig. \ref{fig:dphijj_ps_comparison}, the differential distributions for $\Delta \phi_{jj}$ are shown for the NLO SM matched with \textsc{Pythia8}, with \textsc{Herwig7} and at FO. \textsc{Pythia8} usually applies the global recoil scheme, that is not suitable for VBF processes; a dipole recoil one should be employed instead \cite{Jager:2020}, but this is possible in \textsc{MadGraph5} only at LO, as the counterterms for the shower at NLO are derived assuming the global scheme. In terms of the total fiducial cross section, the best agreement of SM predictions with the ATLAS measurements is obtained when \textsc{Herwig7} is used, but at differential level, FO calculations seem to better agree in the central bins. Since this differential distribution is particularly important for $O_W$, we choose to use the FO generations in our analysis. However, the NLO+PS results for the interference between the SM and the dimension-6 operator show similar $K$-factors. In addition, choosing the generator and shower that more precisely reproduce the data with the SM results might fit away room for new-physics contributions.

The integral and the measurable absolute-valued cross sections are shown in Table \ref{tab:meas} for the linear term, together with the asymmetry measured by $\Delta \phi_{jj}$. These results come from LO event generation, with no PS applied. $\sigma^{|\text{meas}|}$ is computed by summing the interference squared amplitude of each event over all the permutations of initial- and final-state momenta and over all the possible helicity configurations for all the subprocesses; each squared amplitude is weighted by its PDF. It can be seen that a suppression occurs for the interference, as its cross section is much lower than $\sigma^{|\text{int}|}$, but the $\sigma^{|\text{meas}|}$ value shows that most of it can be accessed at experiments and the dijet azimuthal difference can restore a large part of it.

The LO differential distributions for $\Delta \phi_{jj}$ are shown in Fig. \ref{fig:dphijj_lo}, for both the SM and the linear term in $C_W$, with no PS. The positive- and negative-weighted contributions to the interference are also reproduced separately, confirming the ability of this observable to very well separate the opposite-sign components of the interference. The binning we use is $[0,\pi/4,\pi/2,3\pi/4,7\pi/8,15\pi/16,\pi]$ and its symmetric around zero. The NLO predictions for the same distributions at FO are reported in Fig. \ref{fig:dphijj_nlo}, for the SM, $\mathcal{O} (1/\Lambda^2)$ and $\mathcal{O} (1/\Lambda^4)$ contributions. The {\it K}-factors are shown in each bin, with the relative numerical and scale uncertainties, together with the cancellation level $R_{w\pm}$ in \eqref{Rwpm} for the LO interference. The differential {\it K}-factors are stable around $\sim 1$ for this variable, as it happens for the total cross-section ones; indeed, the cancellation level is far from zero in each bin.

We provide 68\% and 95\% Confidence Level (CL) bounds on $C_W/\Lambda^2$ from this variable in Fig. \ref{fig:bounds}. They are obtained by comparing the $\Delta \phi_{jj}$ distributions and the results in \cite{Atlas:2020zjj}, as shown in Table \ref{tab:dataset}. The NLO SM values we use come from the \textsc{Herwig7+Vbfnlo} prediction \cite{Vbfnlo} in the ATLAS paper, while the linear and quadratic corrections are computed at FO. We use the correlation matrix from ATLAS for the numerical uncertainties, while the systematic and scale ones are added in quadrature on the diagonal. It can be seen that the bounds from the interference are comparable to the ones that include the quadratic term as well. Our bounds are of the same order of magnitude than the ones obtained by the ATLAS Collaboration.

\begin{figure}
   \caption{\small{$WZ$ interference cross section per bin at LO without PS, as a function of $p_T^Z$ and $\phi_{WZ}$, in the two cases in which the {\it Z} leptons have helicities $\pm 1$ and $\mp 1$. Red (blue) areas mark where the cross section is positive (negative), as the positive- (negative-) weighted contribution dominates there. The black dashed lines separate the phase-space areas in \eqref{cuts}. The uncertainties are not shown}} \label{fig:pTz_phiWZ_lo}
   \includegraphics[width=.49\textwidth]{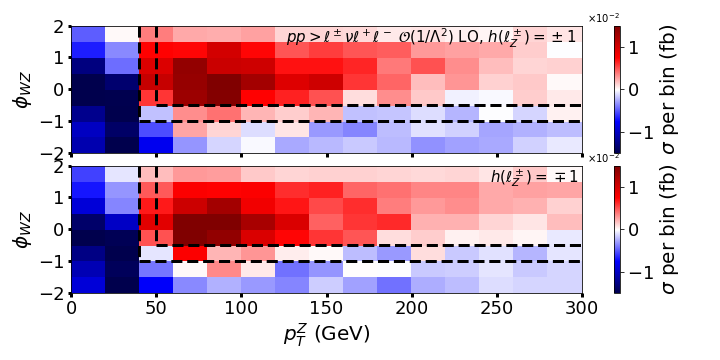}
\end{figure}

\begin{figure}
   \caption{\small{LO and NLO differential cross-section distributions for $\phi_{WZ}$, over all the phase space (\textit{top}) and when specific cuts on $p_T^Z$ and $\phi_{WZ}$ are applied (\textit{centre} and \textit{bottom}). The black (orange, green) line represents the SM, divided by 50 (interference, quadratic correction divided by 4). The {\it K}-factors are also shown, together with their numerical and scale uncertainties. For each case, the relative cancellation for LO interference is plotted. Note the different variable range in the central plot, due to the cuts}} \label{fig:phiWZ}
   \includegraphics[width=0.49\textwidth]{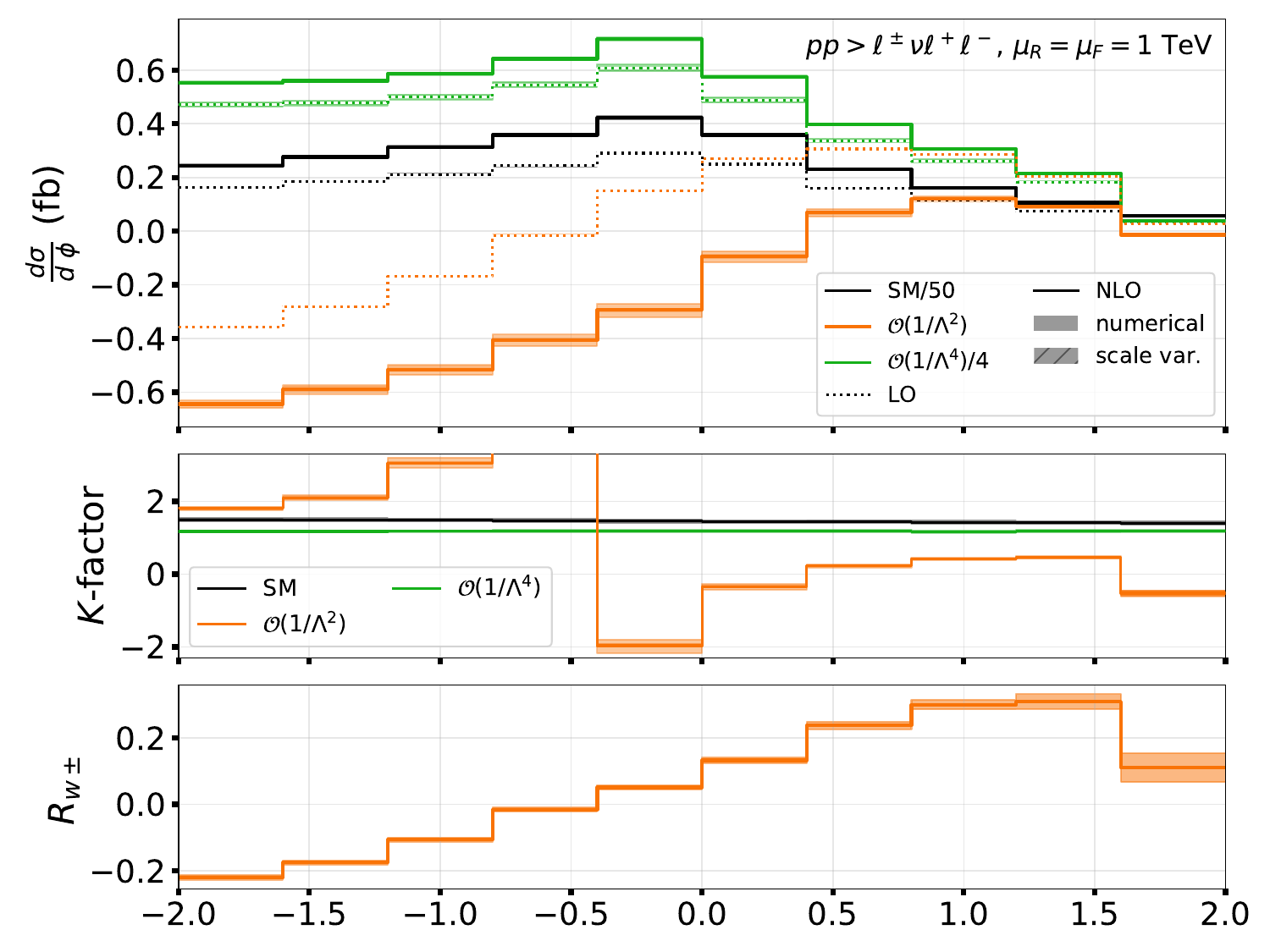}
   \includegraphics[width=0.49\textwidth]{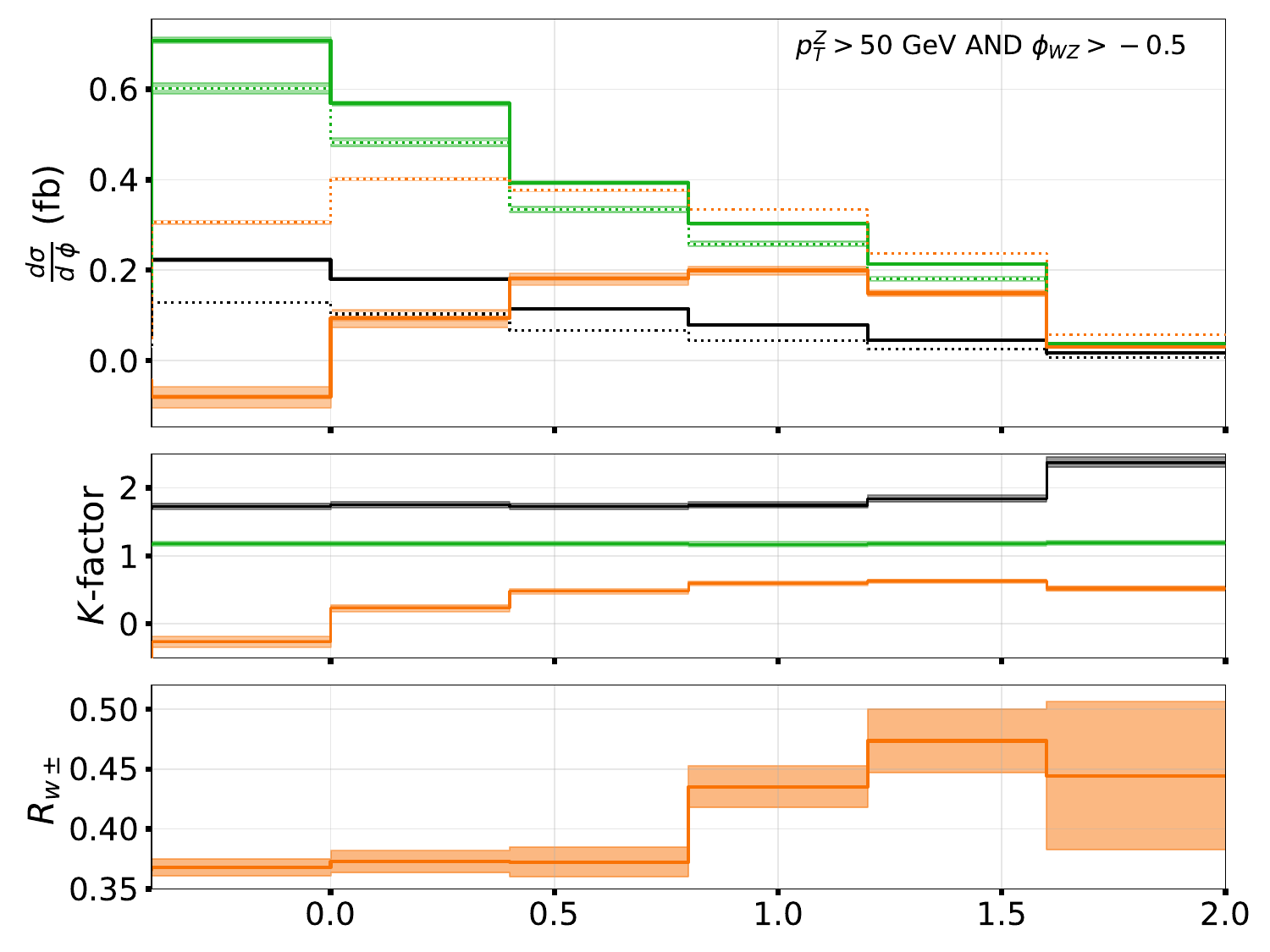}
   \includegraphics[width=0.49\textwidth]{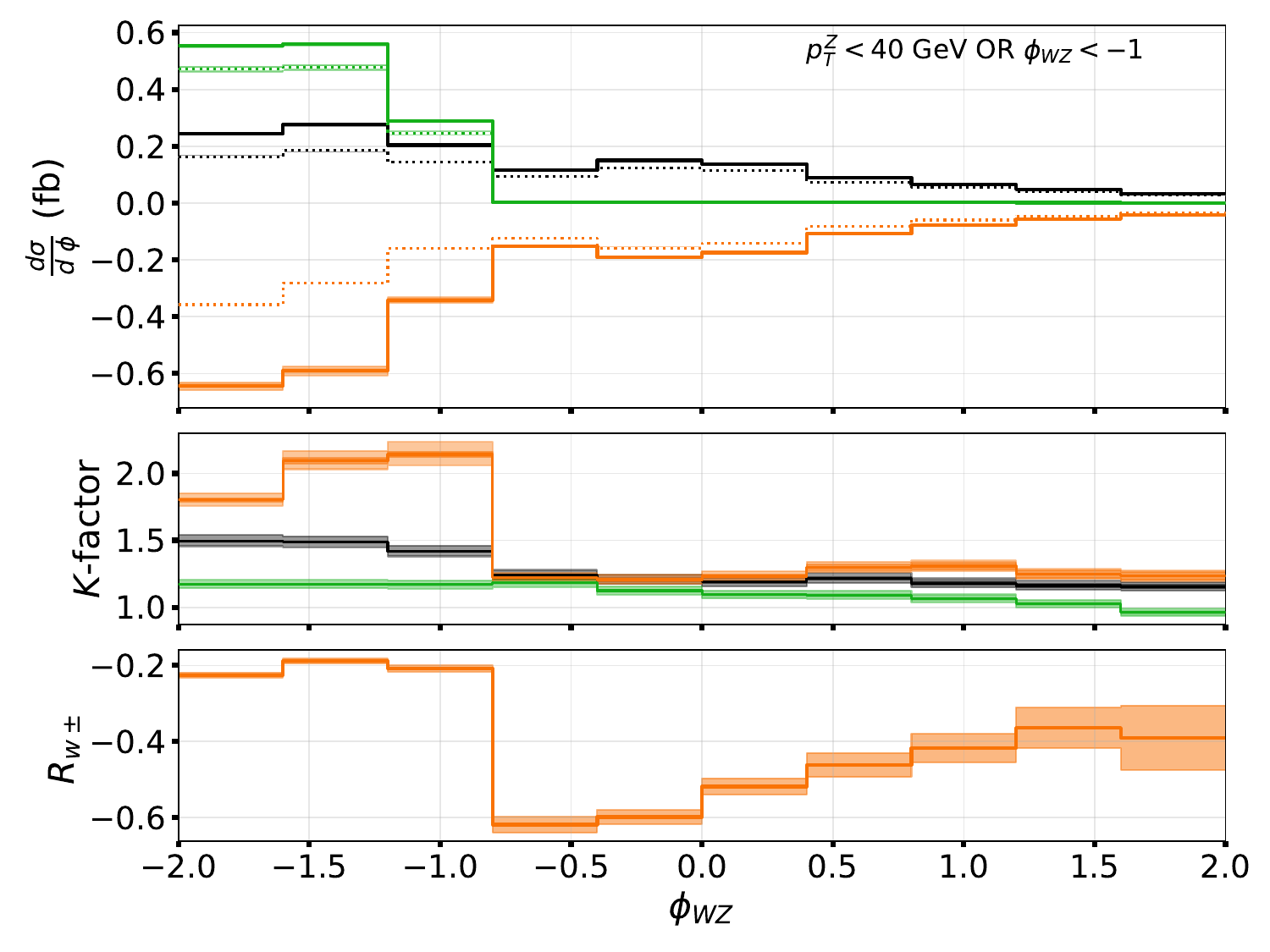}
\end{figure}

\begin{figure}
   \caption{\small{LO SM and interference distributions (without PS) for $\cos \theta^*_{\ell_Z^- Z}$, split according to the $\ell_Z$ helicities. The positive- and negative-weighted contributions to the linear cross-section term are shown in red and blue. The uncertainties are not reported}} \label{fig:cosThZ_lo}
   \includegraphics[width=0.49\textwidth]{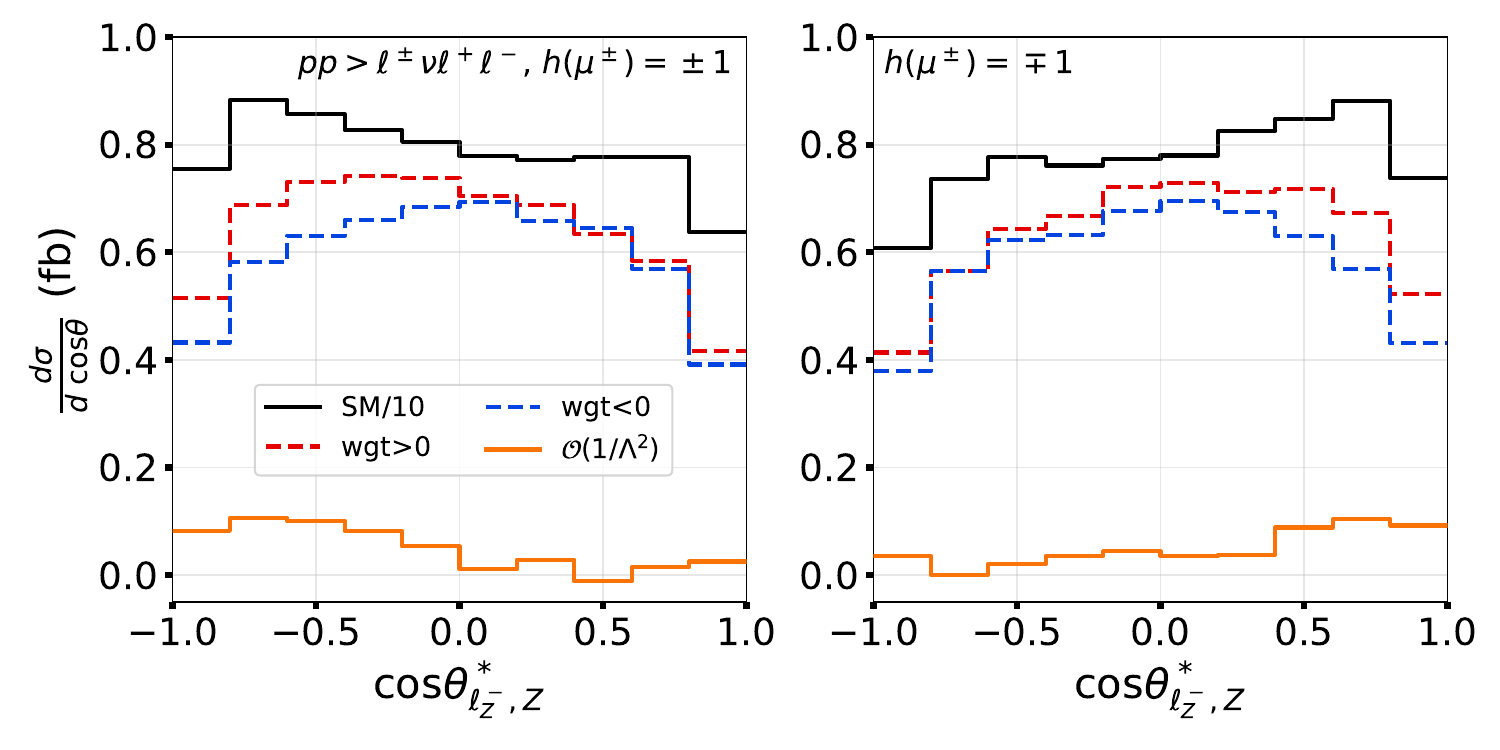}
\end{figure}

\begin{figure}
   \caption{\small{FO LO and NLO differential cross-section distributions for the cosine of the angle between the negatively-charged $\ell_Z$ momentum, in the $Z$-boson rest frame, and the direction of flight of the boson, seen in the $WZ$ CoM frame. The cases of the full phase space (\textit{top}) and when phase-space cuts are applied (\textit{centre} and \textit{bottom}) are shown. The black (orange, green) lines represent the SM, divided by 50 (interference, quadratic term divided by 4). The {\it K}-factors are also shown, together with the relative cancellation}} \label{fig:cosThZ}
   \includegraphics[width=0.49\textwidth]{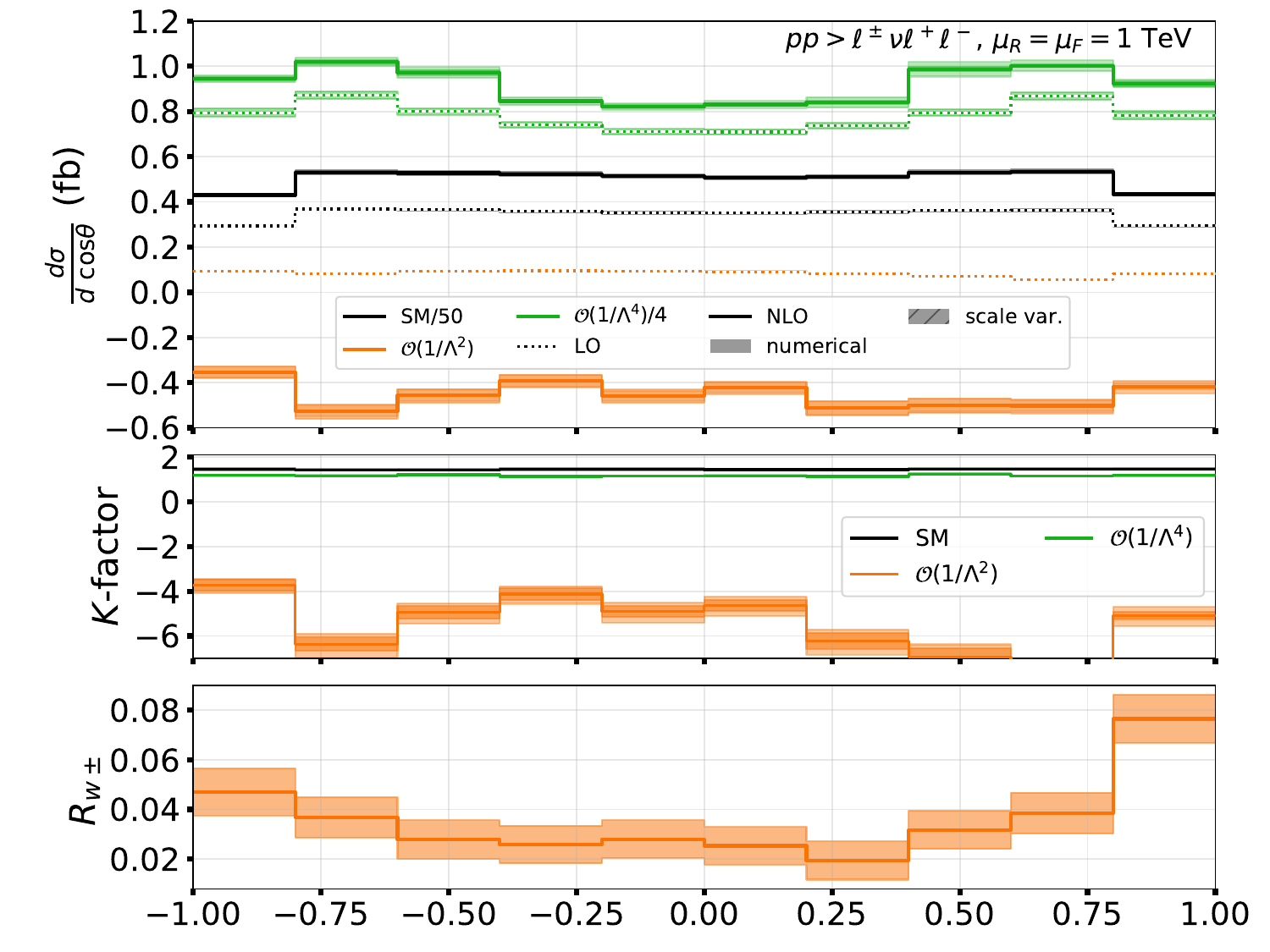}
   \includegraphics[width=0.49\textwidth]{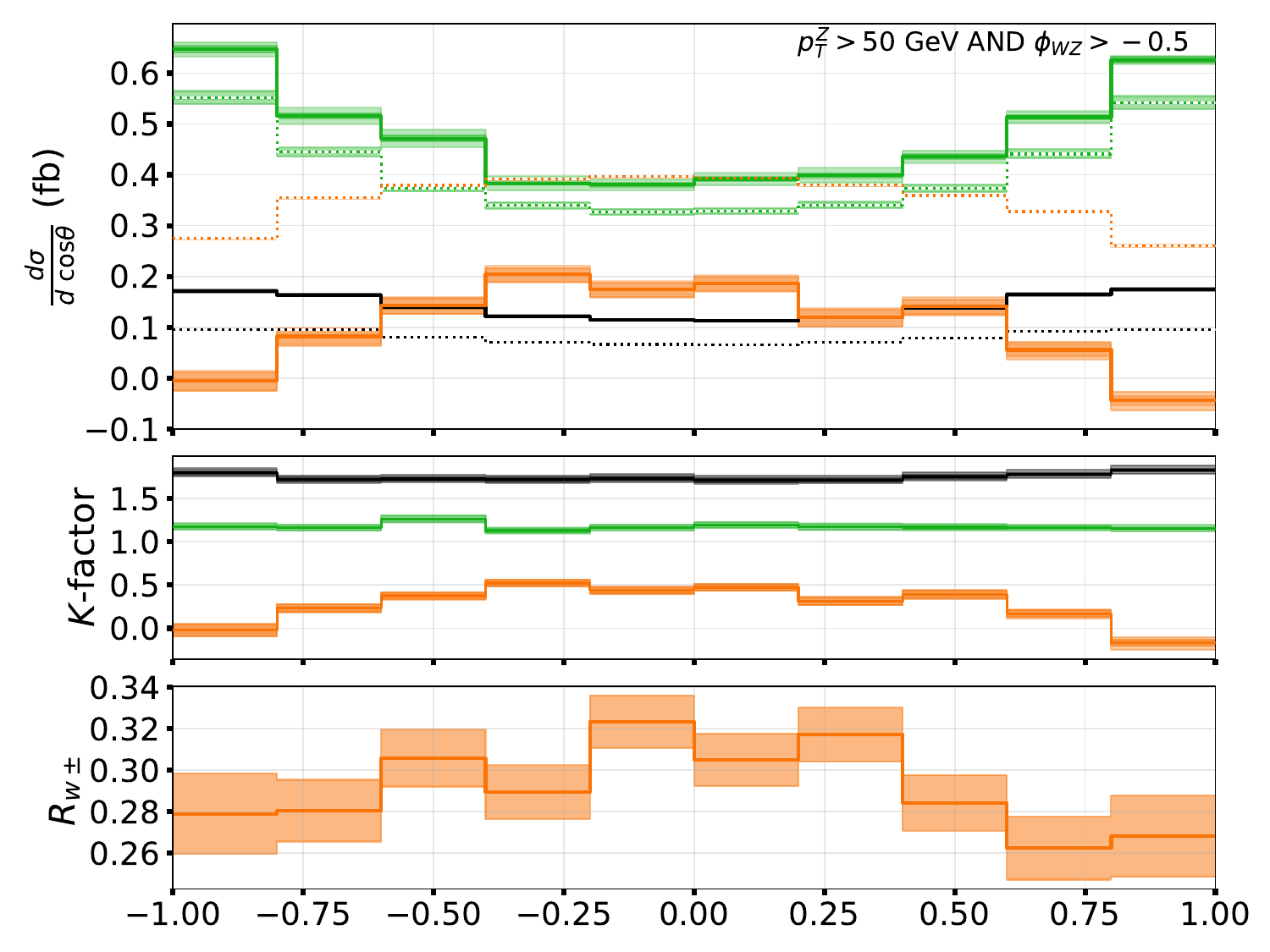}
   \includegraphics[width=0.49\textwidth]{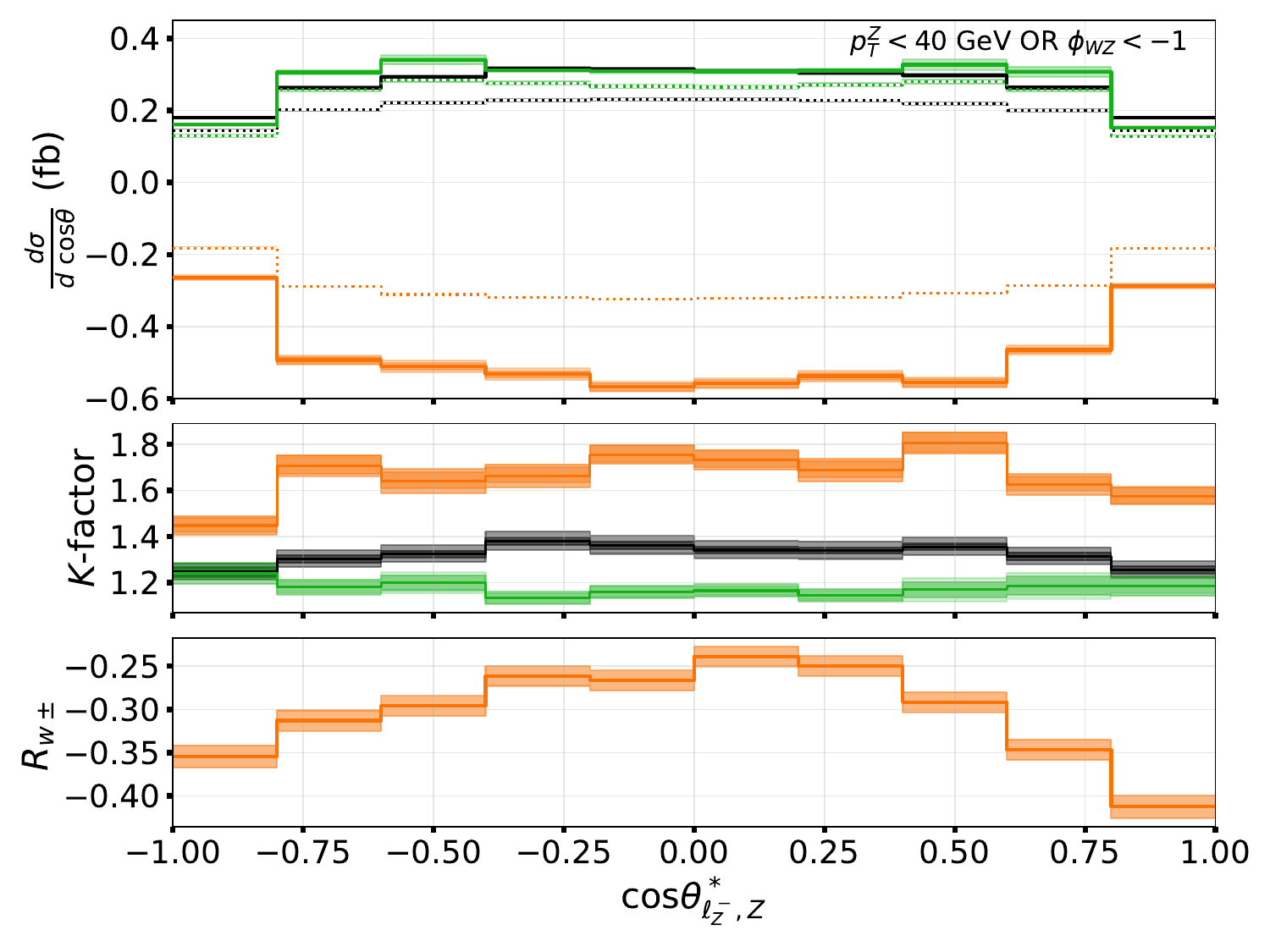}
\end{figure}

\section{Fully-leptonic \textit{WZ} production}

\subparagraph{Calculational details}
We consider the $W^\pm Z$ diboson production, with leptonic decays $W^\pm \rightarrow \ell^\pm \overset{\scriptscriptstyle(-)}{\nu_\ell}$ and $Z \rightarrow \ell^+ \ell^-$, where $\ell= e, \mu$ and the sign refers to their electric charge. A study of the sensitivity to $O_W$ in case of hadronic and semileptonic decays of the $W$ bosons is reviewed in \cite{rafa_will_hadr}, where jet-substructure techniques are employed. In case the three visible leptons are from the same family, the resonant-shape algorithm described in \cite{Atlas:2019wz} is used to assign each of them to their $W$ or $Z$ mother. The phase space is defined according to the criteria given by ATLAS in the same paper: the $p_T$ of $Z$-leptons $\ell_Z$ has to be above 15 GeV, while the $W$-lepton $\ell_W$ one has to be higher than 20 GeV. For all the leptons, we require $|\eta|< 2.5$, and the invariant mass of the $Z$-boson decay products, $M_{\ell_Z^+ \ell_Z^-}$, has to be higher than 81.2 and lower than 101.2 GeV. The angular distance between the two $\ell_Z$ must satisfy $\Delta R > 0.2$, while it has to be greater than 0.3 among $\ell_W$ and the two $\ell_Z$. We furthermore demand the $W$ transverse mass to be above 30 GeV; this variable is defined as $M_T^W=\sqrt{2 p_T^\nu \hspace{1mm} p_T^{\ell_W} \hspace{1mm} (1-\cos{\Delta \phi_{\ell_W \nu}})}$, with $p_T^{\nu,\ell_W}$ the transverse momenta of neutrino and $\ell_W$, and $\Delta \phi_{\ell_W \nu}$ their azimuthal distance.

The main issue in this process is represented by the presence of a neutrino in the final state: since its momentum cannot be measured, we follow the standard strategy of reconstructing it by imposing the $M_{\ell_W \nu}$ invariant mass to be equal to the $W$ pole mass $M_W$ \cite{Riva:2018,Barducci:2019}. From this requirement, up to two possible solutions are obtained, and there is no way to affirm which of them is correct. Previous studies \cite{Rahaman:2020} verified that choosing the one which is smaller in absolute value is more efficient than a completely random extraction, so this is the strategy we adopt. In reality, since the $W$ boson is virtual, its mass is not always equal to $M_W$; even if this criterion provides a good approximation of the true value, there are events in which no real solutions come from it. The neutrino momentum is thus reconstructed by discarding their imaginary part.

To be able to compare with the experimental results in \cite{Atlas:2019wz}, all the cross sections and distributions for this process are averaged over the four different $e^\pm \nu_e e^+ e^-$, $e^\pm \nu_e \mu^+ \mu^-$, $\mu^\pm \nu_\mu \mu^+ \mu^-$, $\mu^\pm \nu_\mu e^+ e^-$ channels; uncertainties are propagated in the average assuming total correlation among them.

\subparagraph{Results}
The LO and NLO total cross sections at FO are shown in Table \ref{tab:xsect}, along with the global {\it K}-factors and their uncertainties. For the SM, the N$^2$LO cross section from \textsc{Matrix} \cite{Grazzini:2017mhc,Gehrmann:2015ora,Denner:2016kdg,Cascioli:2011va,Buccioni:2019sur,Buccioni:2017yxi,Catani:2012qa,Catani:2007vq} is also reported, together with the N$^2$LO/LO ratio: these results are compatible with \cite{Grazzini:2017ckn,Grazzini:2016swo} and the ATLAS ones. The global {\it K}-factor for the interference is large and negative and suggests that the perturbative expansion might not be under control. The integral and measurable absolute-valued cross sections are shown in Table \ref{tab:meas}. $\sigma^{|\text{meas}|}$ is computed by summing the interference squared amplitude of each event over all the permutations of initial- and final-state momenta, over all the possible helicity configurations of all subprocesses and integrating over the longitudinal component of the neutrino momentum; each term is multiplied by its PDF factor. By comparing the actual interference cross section with $\sigma^{|\text{int}|}$, it can be noticed that a large suppression occurs. This is partially lifted when the events where the positively- and negatively-charged $Z$-leptons $\ell_Z^\pm$ have helicities $h=\pm 1$ or $\mp 1$ are considered separately: as it can be seen in the table, the ratios of the measurable over the integral cross sections increase when this distinction is applied. Even if the $\ell_Z$ helicity values are not experimentally accessible, better predictions can come from variables that are sensitive to them.

In \cite{Barducci:2019,NovelTGC:2017}, it is shown that the interference between the $O_W$ operator and the SM in $WZ$ production is proportional to
\begin{equation}
   \phi_{WZ} = \cos (2\phi_W) + \cos (2\phi_Z), \label{phiW}
\end{equation}
where $\phi_V$, $V=W,Z$, is the azimuthal angle between the plane containing the {\it V} boson and the beam axis, and the plane where its decay products lie, in the lab frame. The direction of the latter is defined as the vectorial product of the positive- and negative-helicity lepton three-momenta; since the {\it Z} boson couplings to left- and right-handed leptons are similar, this can be determined only up to an overall sign. This introduces an ambiguity $\phi_Z \longleftrightarrow \phi_Z - \pi$, which however does not affect the value of $\phi_{WZ}$, as it is a function of $\cos (2\phi_Z)$. For what concerns the {\it W} boson, the lepton helicities are constrained by the left-handed nature of the interaction, but another ambiguity is introduced in the longitudinal momentum component of the neutrino. This partially washes away the $2\phi_W$ modulations in the variable above.

The $\mathcal{O} (1/\Lambda^2)$ double-differential distribution for $\phi_{WZ}$ and the transverse momentum of the reconstructed {\it Z} boson, $p_T^Z$, shows similar behaviours when the separation between the events with $h(\ell_Z^\pm)=\pm 1$ and $h(\ell_Z^\pm)=\mp 1$ is applied. The same does not hold for many other observables, and this one presents the largest asymmetry among the ones with this property. This can be seen in Fig. \ref{fig:pTz_phiWZ_lo} for the interference at LO. By cutting the phase space in areas where this double distribution is mainly positive or negative, we can obtain stable and reasonable {\it K}-factors for the variables computed over this process. In particular, the overall positive and negative regions are respectively delimited by
\begin{subequations} \label{cuts}
   \begin{gather}
      p_T^Z> 50 \text{ GeV AND } \phi_{WZ}> -0.5, \label{cuts_a} \\
      p_T^Z< 40 \text{ GeV OR } \phi_{WZ}< -1. \label{cuts_b}
   \end{gather}
\end{subequations}
In the phase-space portion in between these two regions, the LO double distribution changes its sign, yielding unstable and large {\it K}-factors.

The NLO and LO distributions of $\phi_{WZ}$, for the SM, interference and quadratic correction, are shown in Fig. \ref{fig:phiWZ}, with the relative differential {\it K}-factors and cancellation level. In the top plot, that considers the whole phase space, it can be seen that the $\mathcal{O} (1/\Lambda^2)$ \textit{K}-factors jump or become large where the cancellation is almost complete. If the distributions are computed in the phase-space regions defined by the cuts \eqref{cuts}, the \textit{K}-factors are more reasonable: the cancellation $R_{w\pm}$ is always positive (negative) for the first (second) cut, as the opposite-sign weights at LO are partially separated. We use bins of 0.4 from -2 to 2.

For the SMEFT interpretation, the $WZ$ transverse mass was considered in previous studies \cite{ATL-PHYS-PUB-2021-022,Atlas:2019wz}. This variable is defined as
\begin{equation}
   M_T^{WZ} = \sqrt{\left(\sum_\ell p_T^\ell + p_T^\nu \right)^2 - \left(\sum_\ell \vec{p}_T^{\hspace{0.5mm}\ell} + \vec{p}_T^{\hspace{0.5mm}\text{miss}} \right)^2},
\end{equation}
where the sums include the charged leptons. This distribution is useful to probe those SMEFT operators whose effects increase with the centre-of-mass (CoM) energy $\sqrt{s}$. The LO and NLO plots for this observable are presented in Fig. \ref{fig:mTwz}, for the SM, linear and quadratic terms, together with $R_{w\pm}$ and the differential \textit{K}-factors. It can be seen how these are more regular when the cuts \eqref{cuts} over $\phi_{WZ}$ and $p_T^Z$ are applied, showing that less suppression is present in these phase-space regions. In the first bin of the central plot, the $K$-factors are large for all three contributions even though the cancellation level is far from zero, as the cross section in the bin is small. For the whole phase-space case, SM N$^2$LO predictions from \textsc{Matrix} are also plotted, together with the N$^2$LO/LO differential ratio and data from \cite{Atlas:2019wz}. The bins are delimited at $[0,140,180,250,450,600,13000]$ GeV.

Another previous work \cite{Dawson:2019} about SMEFT corrections to $WZ$ production suggests other angular variables that are sensitive to new physics effects. In particular, we focus on $\cos \theta^*_{\ell_Z^- Z}$, the cosine of the angle between the negatively-charged $\ell_Z$ momentum, in the $Z$-boson rest frame, and the direction of flight of the boson, seen in the $WZ$ CoM frame; this coordinate system is defined in \cite{Dixon:2011}. The CoM reference system reconstruction is affected by the uncertainty on the neutrino momentum. The LO SM distribution for this variable is sensitive to the $\ell_Z$ helicities, as it can be seen in Fig. \ref{fig:cosThZ_lo}: the two plots have different trends, with more events close to a variable value of -1 or 1 in the two cases. This dependency, though, is not evident in the $\mathcal{O} (1/\Lambda^2)$ term. The {\it K}-factors for this observable are shown in Fig. \ref{fig:cosThZ}: while they are, at interference level, all negative when considering the whole phase space, they improve when the cuts \eqref{cuts} are applied, as the LO and NLO distributions have same sign in the two regions. We use bins of 0.2 from -1 to 1.

\begin{figure*}
   \caption{\small{68\% and 95\% CL LO ({\it dotted}) and NLO ({\it continuous}) bounds on $C_W /\Lambda^2$, with and without the inclusion of the quadratic term, coming from the processes discussed in this work. For $WZ$, we show the limits over the whole phase space and in the two regions \eqref{cuts}; for $W \gamma$, limits from two double distributions are reported. The variables we use are noted next to the process definitions. The bounds in the gray area come from comparison with the best SM distribution we generated, while the others with real data; a summary of the used datasets can be found in Table \ref{tab:dataset}. The numerical values for the limits are reported on the right}} \label{fig:bounds}
   \includegraphics[width=\textwidth]{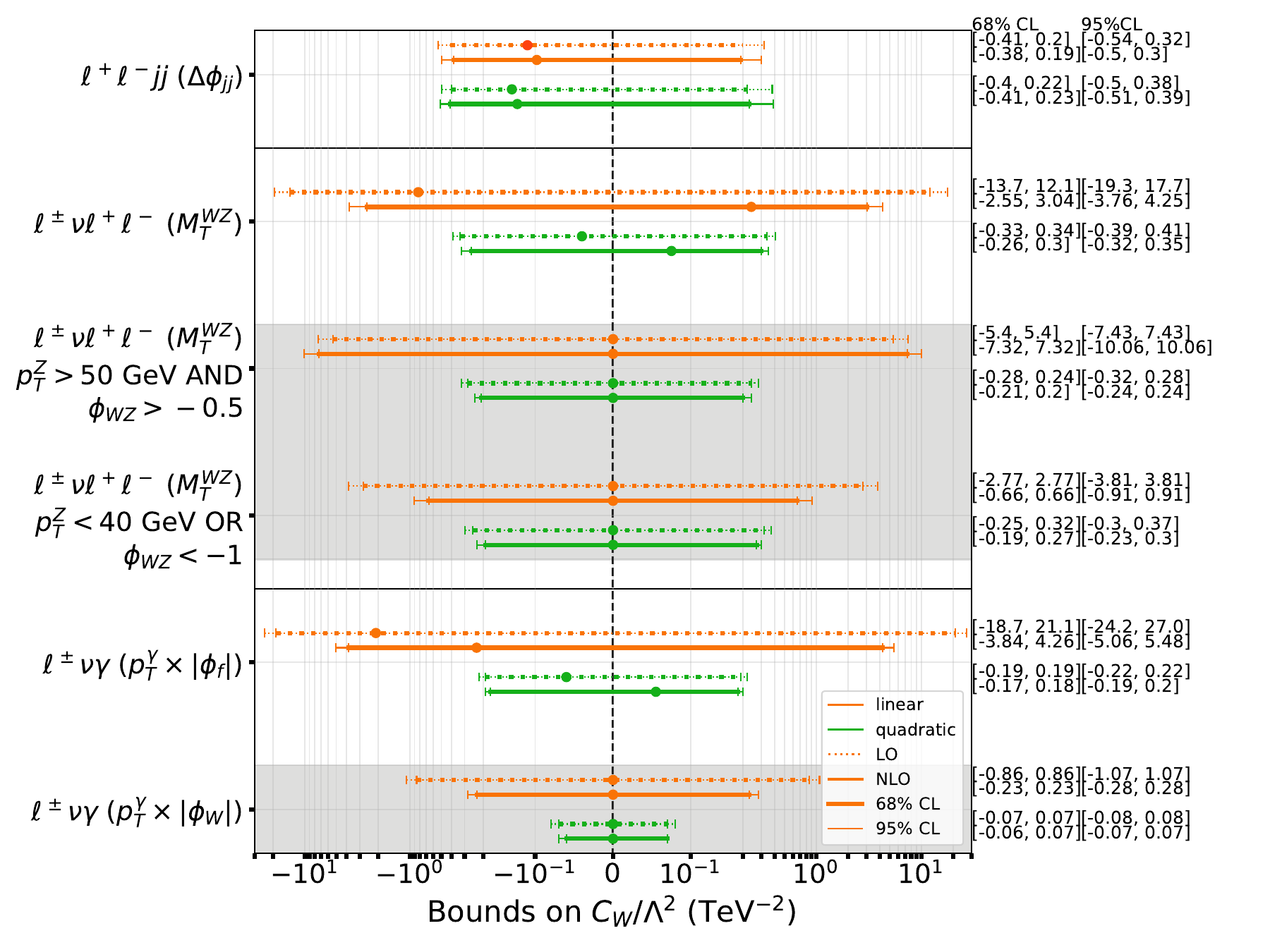}
\end{figure*}

The bounds on $C_W/\Lambda^2$ that we obtain from $M_T^{WZ}$ are shown in Fig. \ref{fig:bounds}, on the whole phase space and in the two regions defined in \eqref{cuts}. As summarised in Table \ref{tab:dataset}, the limits in the first case come from the comparison with the experimental results in \cite{Atlas:2019wz}, using the correlation matrix given there for numerical uncertainties, while the scale variation ones are summed in quadrature on the diagonal. In the special regions, due to the lack of real data, we confront predictions with our SM NLO one. It can be seen that the limits from LO interference improve when the $p_T^Z, \phi_W$ cuts are applied, and that in region \eqref{cuts_a} they are more stringent at LO than NLO due to the {\it K}-factors being smaller than one. The region \eqref{cuts_b} provides constraints that are quite stronger than the full-phase space ones.

\begin{figure}
   \caption{\small{LO and NLO differential cross sections and {\it K}-factors for $\phi_W$, at SM, linear and quadratic orders. The numerical and scale uncertainties are reported for each result. The relative cancellation between positive- and negative-weighted events is plotted for the interference at LO, with numerical error bars}} \label{fig:phiW_fig}
   \includegraphics[width=0.49\textwidth]{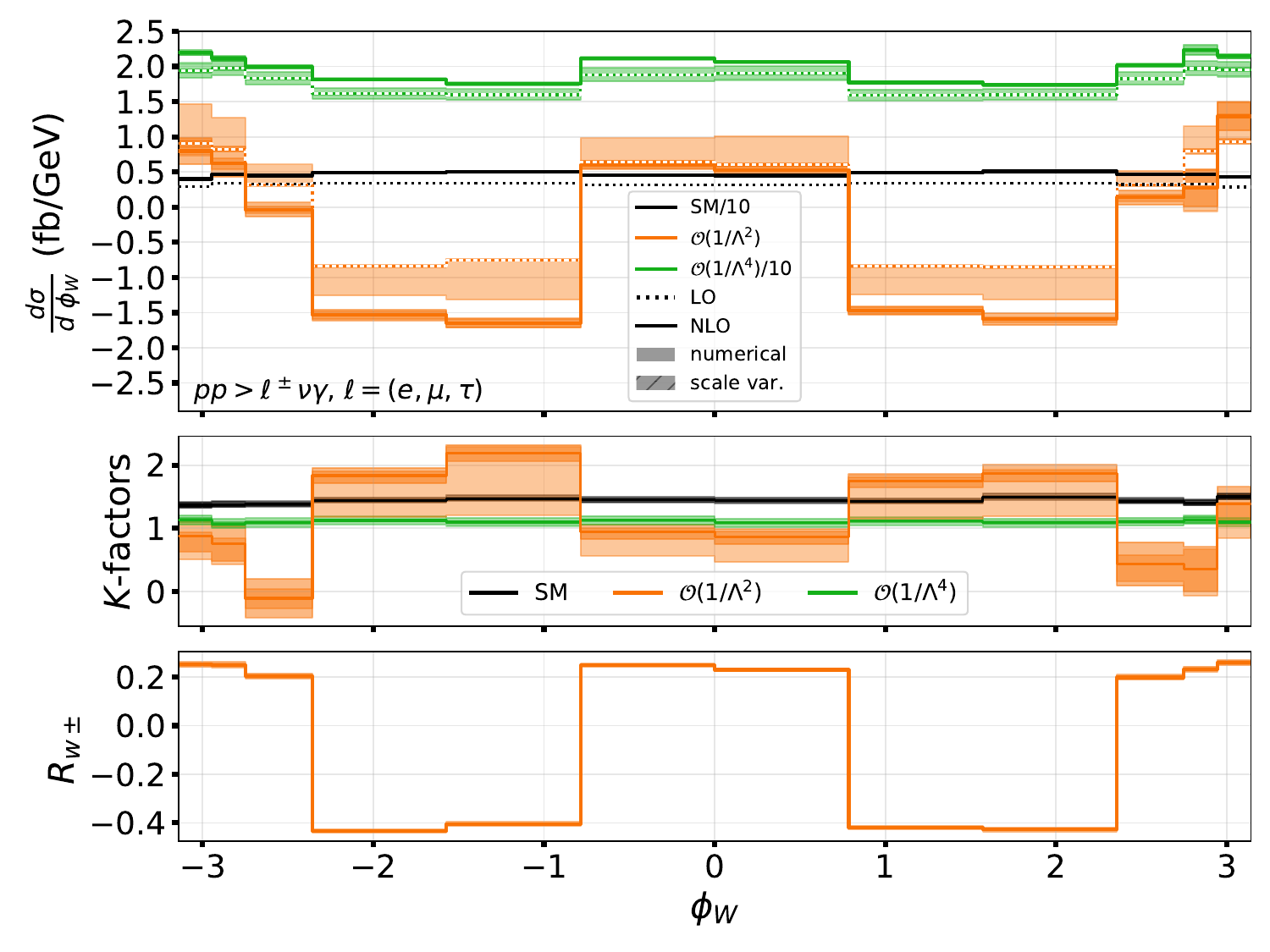}
\end{figure}

\begin{figure}
    \caption{\small{LO and NLO double differential distribution of $p_T^\gamma \times |\phi_f|$. The black (orange, green) line shows the SM divided by 10 (interference, quadratic divided by 10). The blue dots with error bars represent the experimental data, divided by 10. The \textit{K}-factors and the $\mathcal{O}(1/\Lambda^2)$ cancellation are also reported}} \label{fig:pTa_phif}
	\includegraphics[width=0.49\textwidth]{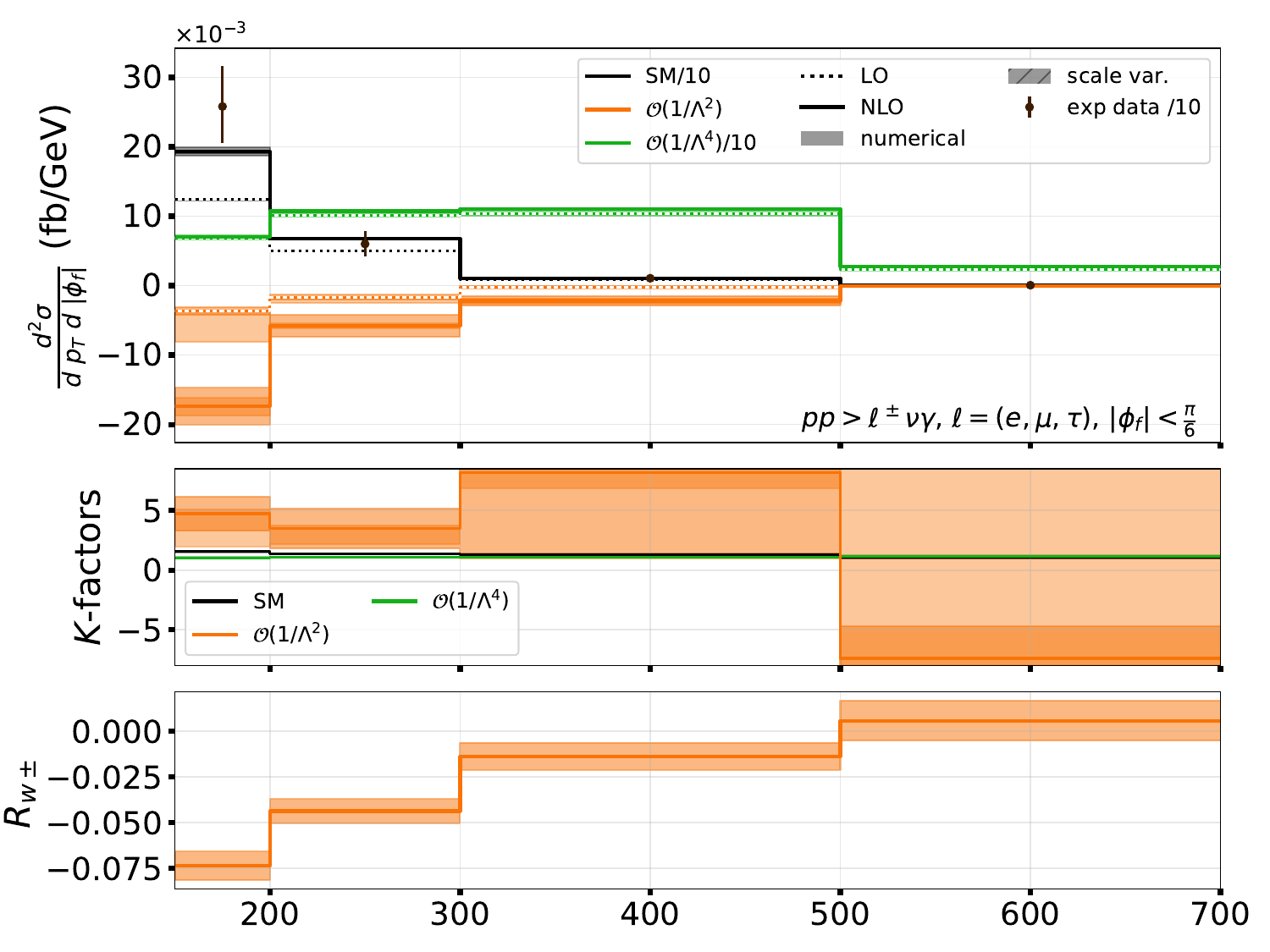}
	\includegraphics[width=0.49\textwidth]{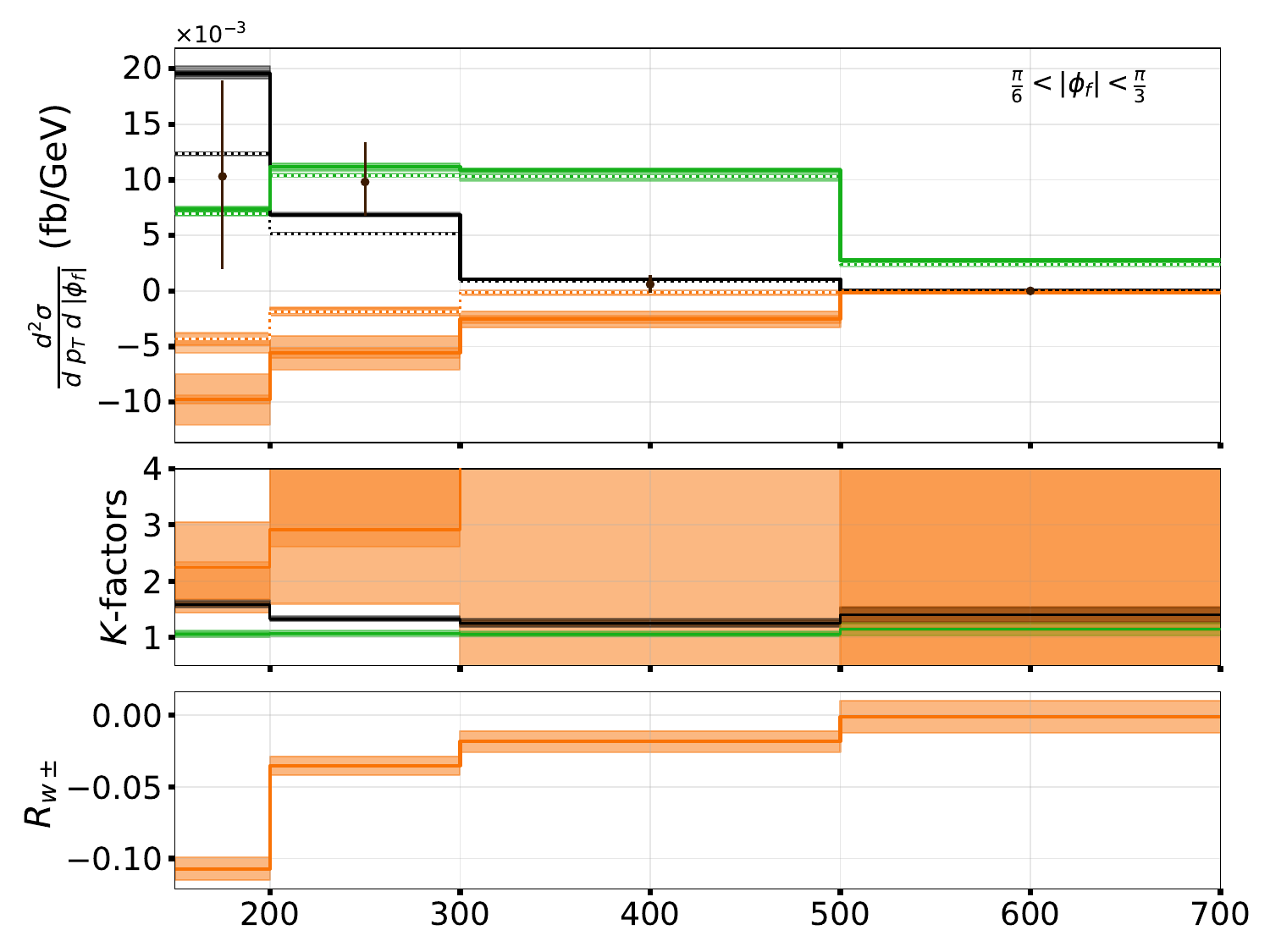}
	\includegraphics[width=0.49\textwidth]{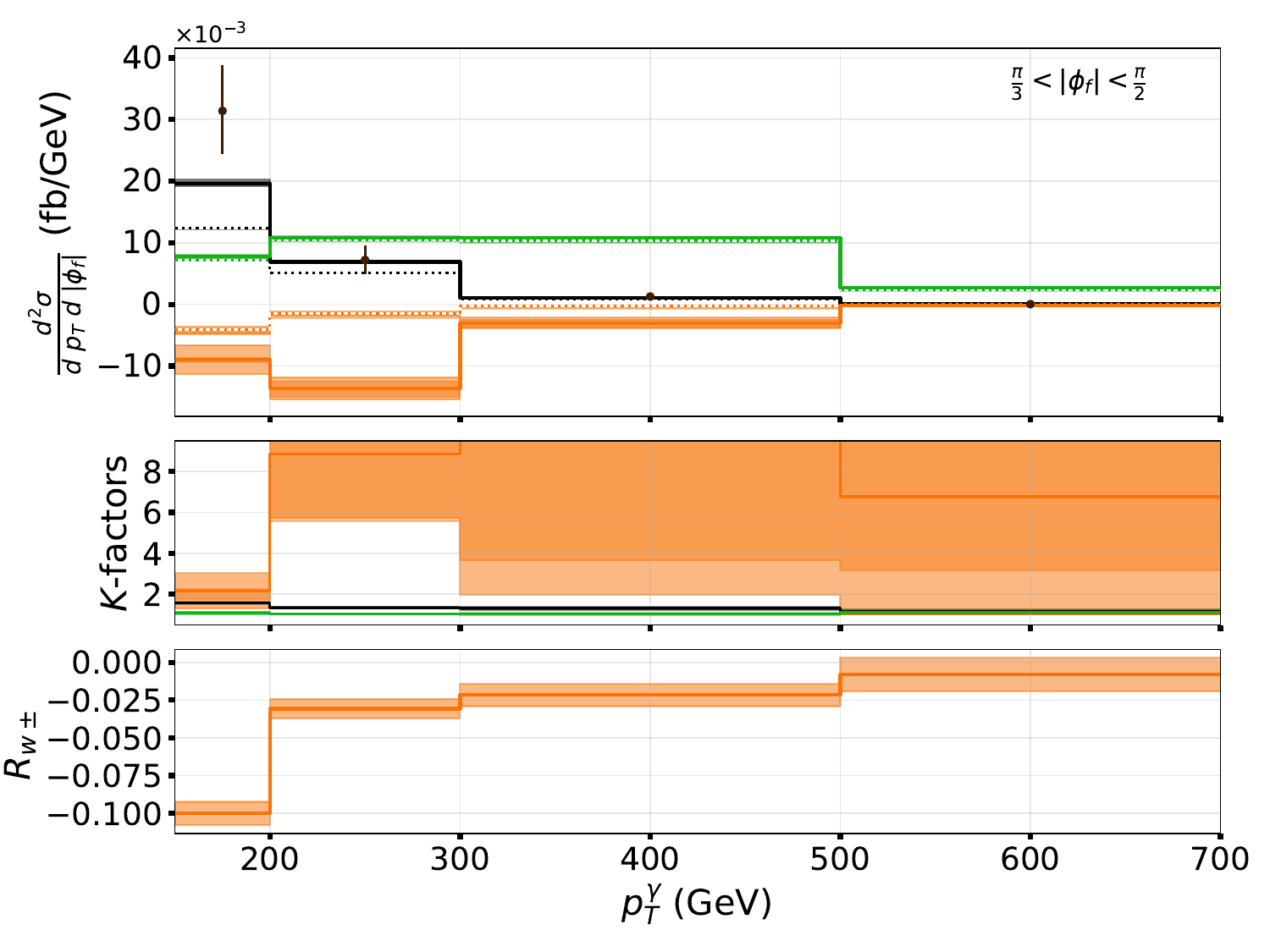}
\end{figure}

\begin{figure}
    \caption{\small{LO and NLO double-differential distribution of $p_T^\gamma \times |\phi_W|$. The black (orange, green) line shows the SM divided by 10 (interference, quadratic divided by 10). The \textit{K}-factors and the $\mathcal{O}(1/\Lambda^2)$ cancellation are also reported}} \label{fig:pTa_phiW}
	\includegraphics[width=0.49\textwidth]{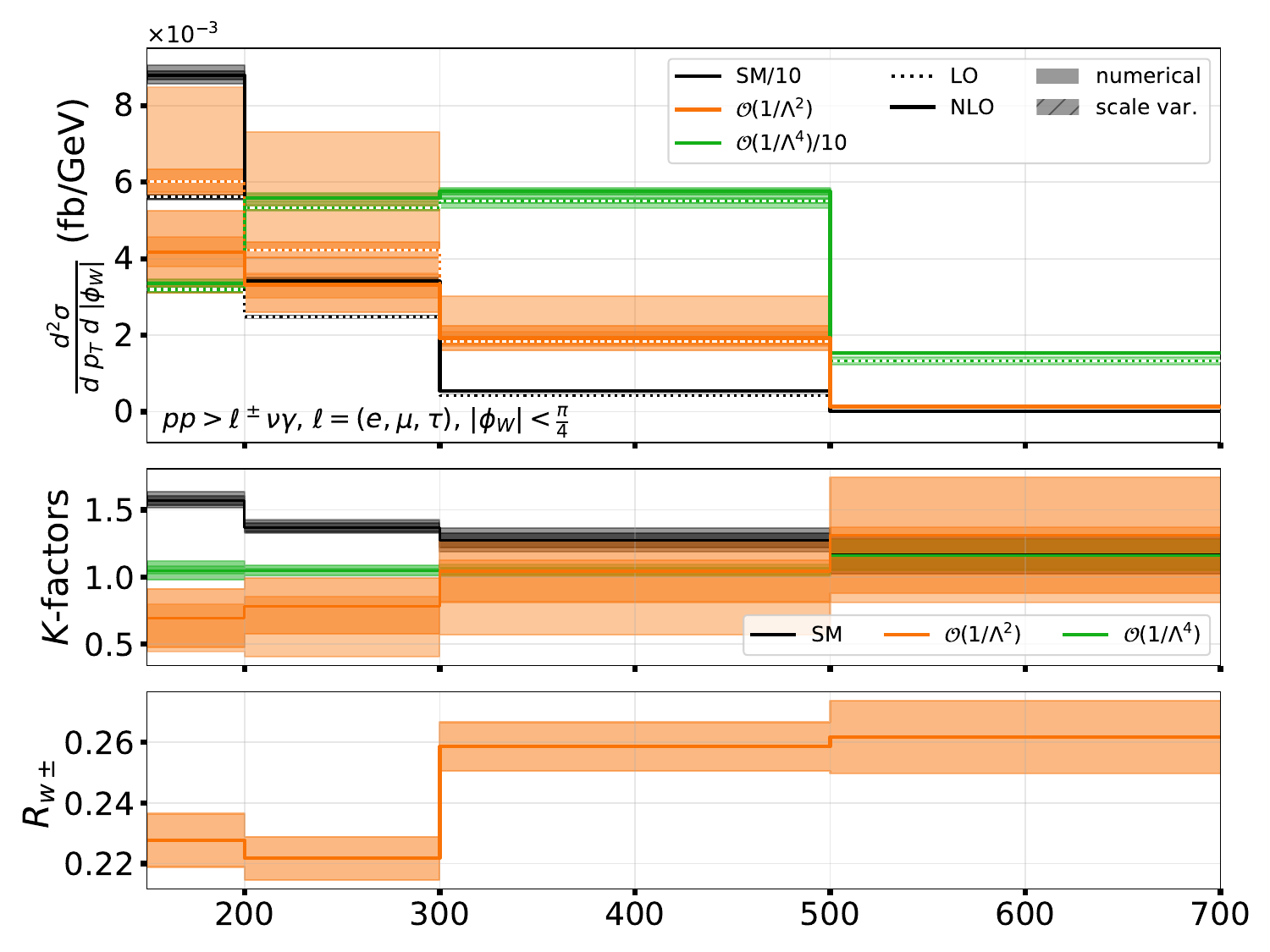}
	\includegraphics[width=0.49\textwidth]{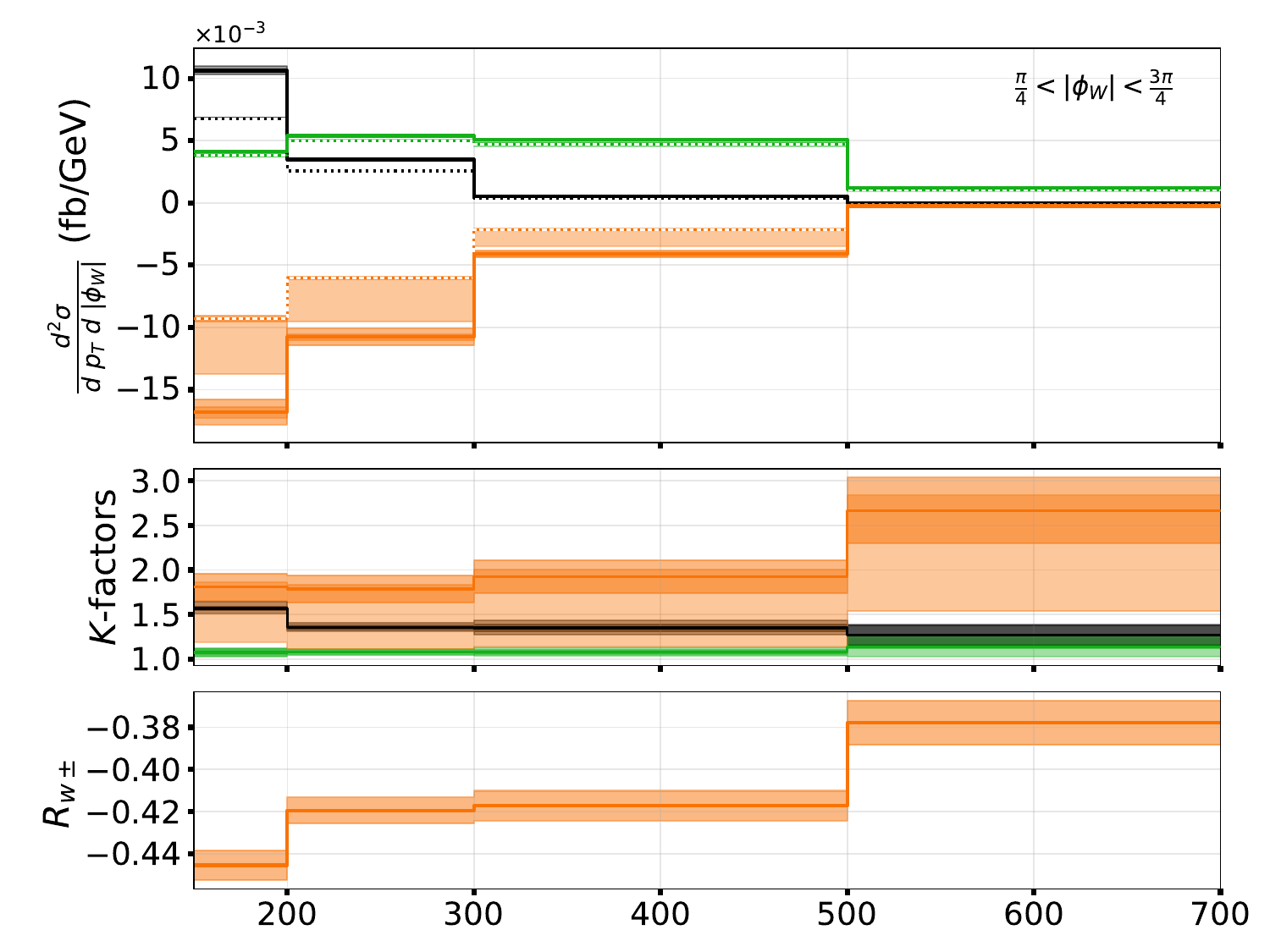}
	\includegraphics[width=0.49\textwidth]{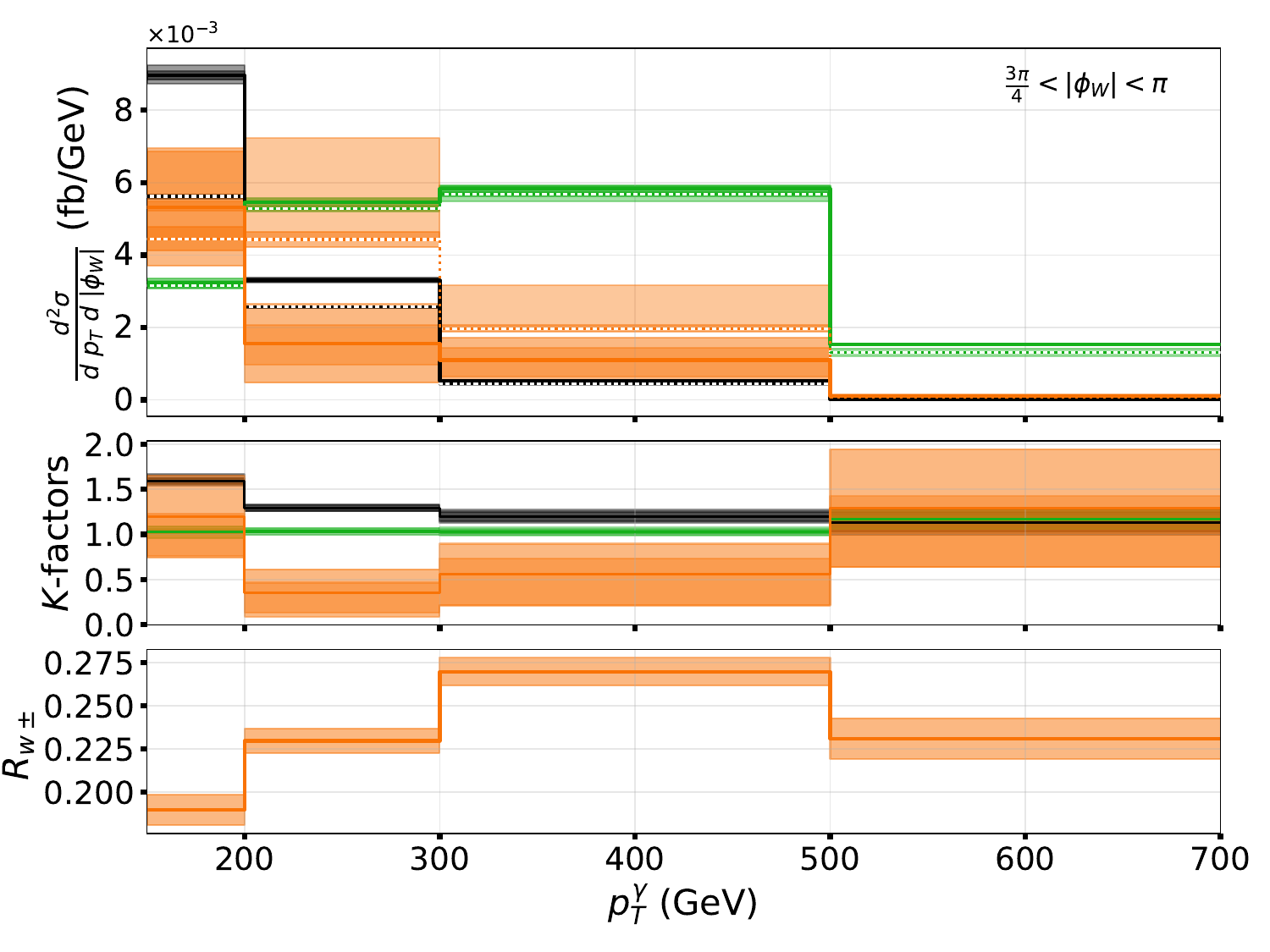}
\end{figure}

\begin{table*}
\caption{\small{Summary of the experimental measurements used to obtain bounds on $C_W/\Lambda^2$, for each process investigated in this paper. In the signal regions \eqref{cuts} and for the second $W\gamma$ observable, that have not been measured yet, the experimental distributions are considered to follow the NLO SM ones}} \label{tab:dataset}
\begin{tabular}{cc|cccc}
   \hline
   Process & Observable & $\sqrt{s}$, $\mathcal{L}$ & Final state & $N_\text{data}$ & Ref. \\ \hline \hline
   $Zjj$ & $d\sigma/d \Delta \phi_{jj}$ & 13 TeV, 139 fb${}^{-1}$ & $\ell^+ \ell^- +$jets, $\ell=e,\mu$ & 12 & \cite{Atlas:2020zjj} \\ \hline
   $W^\pm Z$, full phase space & $d\sigma/d M_T^{WZ}$ & 13 TeV, 36.1 fb${}^{-1}$ & $\ell^\pm \nu \ell^+ \ell^-$, $\ell=e,\mu$ & 6 & \cite{Atlas:2019wz} \\
   \makecell{$W^\pm Z$, $p_T^Z > 50$ GeV \\ AND $\phi_{WZ}>-0.5$} & $d\sigma/d M_T^{WZ}$ & \multicolumn{4}{c}{Exp. data taken as NLO SM} \\
   \makecell{$W^\pm Z$, $p_T^Z < 40$ GeV \\ OR $\phi_{WZ}<-1$} & $d\sigma/d M_T^{WZ}$ & \multicolumn{4}{c}{Exp. data taken as NLO SM} \\ \hline
   $W^\pm \gamma$ & $d^2 \sigma /(d p_T^\gamma \hspace{1mm} d|\phi_f|)$ & 13 TeV, 138 fb${}^{-1}$ & $\ell^\pm \nu \gamma$, $\ell=e,\mu,\tau$ & 12 & \cite{WgCMS:2021}\\
   $W^\pm \gamma$ & $d^2 \sigma /(d p_T^\gamma \hspace{1mm} d|\phi_W|)$ & \multicolumn{4}{c}{Exp. data taken as NLO SM} \\ \hline
\end{tabular}
\end{table*}

\section{Leptonic \textit{W} + photon production}

\subparagraph{Calculational details}
We present some predictions for the $W^\pm \gamma$ production, with the $W$ boson decaying as $W^\pm \rightarrow \ell^\pm \overset{\scriptscriptstyle(-)}{\nu_\ell}$. All three lepton families are considered: $\ell= e, \mu, \tau$, with the $\tau$ decays handled by {\sc Pythia8}. The phase space is delimited as in the EFT analysis in \cite{WgCMS:2021} by the CMS Collaboration: a minimum $p_T$ of 80 GeV and a maximum $|\eta|$ of 2.5 are required for the lepton, while for the missing transverse energy we demand $p_T^\text{miss} > 40$ GeV. For the photon, requirements include $p_T > 150$ GeV, $|\eta| < 2.5$, $\Delta R_{\ell \gamma} > 0.7$. No jets with $p_T > 30$ GeV and $|\eta| < 2.5$ are allowed. 

\subparagraph{Results}
The total cross sections, at LO and NLO, and the {\it K}-factors are summarised in Table \ref{tab:xsect}, together with the numerical and scale uncertainties: they come from NLO event generation matched with \textsc{Pythia8}. The global {\it K}-factor for the interference is much larger than one. Table \ref{tab:meas} reports the measurable and integral cross sections: $\sigma^{|\text{meas}|}$ is computed as in $WZ$ production, by summing the interference squared amplitude of each event over all the permutations of initial- and final-state momenta, over all the possible helicity configurations for all the subprocesses and integrating over the longitudinal component of the neutrino momentum; all the amplitudes are multiplied by their PDF factor. The comparison with $\sigma^{1/\Lambda^2}$ reveals a large suppression at LO. The significant difference among the interference cross section results in the two tables is a PS effect, as many events gain a jet after it and need to be discarded for the above-mentioned requirements. As in the \textit{Zjj} case, an observable is able to lift the interference cancellation considerably: this is the azimuthal angle $\phi_W$ between the plane with the $W$ boson and the beam axis, and the plane where the decay products lie, in lab frame; it is analogous to the ones defined in \eqref{phiW}. The neutrino-momentum reconstruction, carried on as for $WZ$, introduces an ambiguity in this observable value; this does not prevent us from observing the interference effects \cite{Barducci:2019}. The LO, NLO and {\it K}-factor distributions for this variable are shown in Fig. \ref{fig:phiW_fig} for the SM, interference and quadratic terms, together with the cancellation level. The binning is $[0,\pi/4,\pi/2,3\pi/4,7\pi/8,15\pi/16,\pi]$ and its symmetric around 0.

In the CMS paper, the angle $\phi$ is introduced in the CoM frame as the $\ell^\pm$ azimuthal angle, with the $\hat{z}$ axis along the $W^\pm$ three-momentum and $\hat{y} = \hat{r} \times \hat{z}$, where $\hat{r}$ denotes the Lorentz-boost direction from the lab frame. Given the ambiguity in $\phi$ due to the neutrino reconstruction, the angle $\phi_f$ is employed, with $\phi_f = \phi$ if $|\phi|<\pi/2$, or $\phi_f=\pi-\phi$ if $\phi>\pi/2$ and $\phi_f = -(\pi+\phi)$ if $\phi<-\pi/2$. The EFT analysis is carried on through the double distribution of $p_T^\gamma$ and $|\phi_f|$, with three bins of $\pi/6$ from 0 to $\pi/2$ for the angle, and $[150,200,300,500,800,1500]$ GeV for the photon transverse momentum. The LO and NLO distributions of $p_T^\gamma$, split according to the $|\phi_f|$ interval and with their \textit{K}-factors and $R_{w\pm}$, are shown in Fig. \ref{fig:pTa_phif}, for the SM, interference and quadratic corrections.

It can be seen from Table \ref{tab:meas} that the $p_T^\gamma \times |\phi_f|$ distribution at LO is as suppressed as the total interference cross section. As a comparison, we show the LO and NLO double distribution for $p_T^\gamma \times |\phi_W|$, which is sensible to a larger fraction of $\sigma^{|\text{meas}|}$, in Fig. \ref{fig:pTa_phiW}; we use the same binning for the transverse momentum, and [0,$\pi/4$,$3\pi/4$,$\pi$] for the absolute value of the angle. The \textit{K}-factors and their uncertainties are more stable and reasonable in these plots, as the LO and NLO interference distributions have the same sign in the three regions delimited by $|\phi_W|$, positive in the external ones and negative in the central.

The bounds on $C_W/\Lambda^2$ from the two double distributions are shown in Fig. \ref{fig:bounds}. As reported in Table \ref{tab:dataset}, the limits from the $p_T^\gamma \times |\phi_W|$ variable are obtained out of comparison with the SM NLO distribution we generated, while the bounds for the $p_T^\gamma \times |\phi_f|$ double distribution come from confrontation with the data in \cite{WgCMS:2021}. We assume no correlation among different bins for the $\phi_W$ case, as no matrix is available; for $\phi_f$, the matrix given in the CMS paper is used for numerical uncertainties, while scale variations are added in quadrature on the diagonal. The bounds we obtain from $p_T^\gamma \times |\phi_W|$ are better than the ones from $p_T^\gamma \times |\phi_f|$ already at LO for both the interference and the quadratic level.

\section{Conclusions}
In this paper, we showed results for three process that are affected by the $O_W$ operator, for which the linear correction to the SM is suppressed: EW \textit{Zjj} VBF, \textit{WZ} and $W \gamma$ production. We reported differential distributions for relevant variables, at LO and NLO, for the SM, the linear and the quadratic terms. We showed how the choice of these variables is important to improve the predictions when the linear term is suppressed, and how the \textit{K}-factors display a perturbative expansion under better control and with lower uncertainties when the opposite-sign contributions of the interference at LO are separated with suitable phase-space cuts.

We also presented bounds on $C_W/\Lambda^2$ from the comparison among our predictions and real data. In the $Zjj$ case, the limits that we obtained at linear level are comparable with the ones from the quadratic order, which does not experience the same cancellation. For the other processes, they still are at least three times larger, but of the same order of magnitude if they come from suitable observables, as it can be seen in the $W \gamma$ case. It is also important to remind that the $\mathcal{O}(1/\Lambda^4)$ order in this study is missing the inerference of dimension-8 operators and the SM to be complete. One of the reasons behind the worst bounds from the diboson processes can be traced back to the presence of a neutrino in the final state, which translates in a lower ratio of $\sigma^{|\text{meas}|}$ over $\sigma^{|\text{int}|}$. In principle, the same processes can be analysed with the $W$ decaying hadronically, but a larger background is observed at the LHC in that case. In the VBF case, on the other side, a better reconstruction of the outgoing particles is possible at experimental level.

The variables we propose are fully generic, since they only depend on the kinematics, and can be employed even when the EFT validity breaks down; the corresponding predictions can also be updated at any order. The techniques introduced in this paper can be used in conjunction with machine learning to get better theoretical results for the interference among the SM and SMEFT operators. We showed that, in general, a low cancellation over the phase space is needed to obtain meaningful predictions for the interference with respect to higher-order terms. Although we focused here on QCD corrections, we expect this conclusion to be valid for EW ones as well. In particular, the very good revival obtained in VBF through the $\Delta \phi_{jj}$ distribution suggests that, even if the EW corrections are expected to be large for the interference, they should not yield \textit{K}-factors too far from unity. The same statement cannot be affirmed with the same confidence in the diboson cases, as the unmeasurable cancellation is larger. However, $W \gamma$ displays a strong constraining power, similar to VBF, when interference-reviving distributions are used.

\subparagraph{Acknowledgments}
Computational resources have been provided by the Consortium des Équipements de Calcul Intensif (CÉCI), funded by the Fonds de la Recherche Scientifique de Belgique (F.R.S.-FNRS) under Grant No. 2.5020.11 and by the Walloon Region. M.M. is a Research Fellow of the F.R.S.-FNRS, through the grant \textit{Aspirant}. We are thankful to F. Maltoni, E. Vryonidou, G. Durieux, K. Mimasu, A. Pilkington, N. Clarke Hall and A. Gilbert for discussion and very useful suggestions during this work.

\bibliography{refs.bib}

\begin{thebibliography}{49}%
\makeatletter
\providecommand \@ifxundefined [1]{%
 \@ifx{#1\undefined}
}%
\providecommand \@ifnum [1]{%
 \ifnum #1\expandafter \@firstoftwo
 \else \expandafter \@secondoftwo
 \fi
}%
\providecommand \@ifx [1]{%
 \ifx #1\expandafter \@firstoftwo
 \else \expandafter \@secondoftwo
 \fi
}%
\providecommand \natexlab [1]{#1}%
\providecommand \enquote  [1]{``#1''}%
\providecommand \bibnamefont  [1]{#1}%
\providecommand \bibfnamefont [1]{#1}%
\providecommand \citenamefont [1]{#1}%
\providecommand \href@noop [0]{\@secondoftwo}%
\providecommand \href [0]{\begingroup \@sanitize@url \@href}%
\providecommand \@href[1]{\@@startlink{#1}\@@href}%
\providecommand \@@href[1]{\endgroup#1\@@endlink}%
\providecommand \@sanitize@url [0]{\catcode `\\12\catcode `\$12\catcode
  `\&12\catcode `\#12\catcode `\^12\catcode `\_12\catcode `\%12\relax}%
\providecommand \@@startlink[1]{}%
\providecommand \@@endlink[0]{}%
\providecommand \url  [0]{\begingroup\@sanitize@url \@url }%
\providecommand \@url [1]{\endgroup\@href {#1}{\urlprefix }}%
\providecommand \urlprefix  [0]{URL }%
\providecommand \Eprint [0]{\href }%
\providecommand \doibase [0]{http://dx.doi.org/}%
\providecommand \selectlanguage [0]{\@gobble}%
\providecommand \bibinfo  [0]{\@secondoftwo}%
\providecommand \bibfield  [0]{\@secondoftwo}%
\providecommand \translation [1]{[#1]}%
\providecommand \BibitemOpen [0]{}%
\providecommand \bibitemStop [0]{}%
\providecommand \bibitemNoStop [0]{.\EOS\space}%
\providecommand \EOS [0]{\spacefactor3000\relax}%
\providecommand \BibitemShut  [1]{\csname bibitem#1\endcsname}%
\let\auto@bib@innerbib\@empty
\bibitem [{\citenamefont {Grzadkowski}\ \emph {et~al.}(2010)\citenamefont
  {Grzadkowski}, \citenamefont {Iskrzynski}, \citenamefont {Misiak},\ and\
  \citenamefont {Rosiek}}]{Grzadkowski:2010es}%
  \BibitemOpen
  \bibfield  {author} {\bibinfo {author} {\bibfnamefont {B.}~\bibnamefont
  {Grzadkowski}}, \bibinfo {author} {\bibfnamefont {M.}~\bibnamefont
  {Iskrzynski}}, \bibinfo {author} {\bibfnamefont {M.}~\bibnamefont {Misiak}},
  \ and\ \bibinfo {author} {\bibfnamefont {J.}~\bibnamefont {Rosiek}},\
  }\bibfield  {title} {\enquote {\bibinfo {title} {{Dimension-Six Terms in the
  Standard Model Lagrangian}},}\ }\href {\doibase 10.1007/JHEP10(2010)085}
  {\bibfield  {journal} {\bibinfo  {journal} {JHEP}\ }\textbf {\bibinfo
  {volume} {10}},\ \bibinfo {pages} {085} (\bibinfo {year} {2010})},\ \Eprint
  {http://arxiv.org/abs/1008.4884} {arXiv:1008.4884 [hep-ph]} \BibitemShut
  {NoStop}%
\bibitem [{\citenamefont {Buchmuller}\ and\ \citenamefont
  {Wyler}(1986)}]{Buchmuller:1985jz}%
  \BibitemOpen
  \bibfield  {author} {\bibinfo {author} {\bibfnamefont {W.}~\bibnamefont
  {Buchmuller}}\ and\ \bibinfo {author} {\bibfnamefont {D.}~\bibnamefont
  {Wyler}},\ }\bibfield  {title} {\enquote {\bibinfo {title} {{Effective
  Lagrangian Analysis of New Interactions and Flavor Conservation}},}\ }\href
  {\doibase 10.1016/0550-3213(86)90262-2} {\bibfield  {journal} {\bibinfo
  {journal} {Nucl. Phys. B}\ }\textbf {\bibinfo {volume} {268}},\ \bibinfo
  {pages} {621--653} (\bibinfo {year} {1986})}\BibitemShut {NoStop}%
\bibitem [{\citenamefont {Azatov}\ \emph
  {et~al.}(2017{\natexlab{a}})\citenamefont {Azatov}, \citenamefont {Contino},
  \citenamefont {Machado},\ and\ \citenamefont {Riva}}]{Azatov:2016sqh}%
  \BibitemOpen
  \bibfield  {author} {\bibinfo {author} {\bibfnamefont {A.}~\bibnamefont
  {Azatov}}, \bibinfo {author} {\bibfnamefont {R.}~\bibnamefont {Contino}},
  \bibinfo {author} {\bibfnamefont {C.S.}\ \bibnamefont {Machado}}, \ and\
  \bibinfo {author} {\bibfnamefont {F.}~\bibnamefont {Riva}},\ }\bibfield
  {title} {\enquote {\bibinfo {title} {{Helicity selection rules and
  noninterference for BSM amplitudes}},}\ }\href {\doibase
  10.1103/PhysRevD.95.065014} {\bibfield  {journal} {\bibinfo  {journal} {Phys.
  Rev. D}\ }\textbf {\bibinfo {volume} {95}},\ \bibinfo {pages} {065014}
  (\bibinfo {year} {2017}{\natexlab{a}})},\ \Eprint
  {http://arxiv.org/abs/1607.05236} {arXiv:1607.05236 [hep-ph]} \BibitemShut
  {NoStop}%
\bibitem [{\citenamefont {Dixon}\ and\ \citenamefont
  {Shadmi}(1994)}]{Dixon:1994}%
  \BibitemOpen
  \bibfield  {author} {\bibinfo {author} {\bibfnamefont {L.}~\bibnamefont
  {Dixon}}\ and\ \bibinfo {author} {\bibfnamefont {Y.}~\bibnamefont {Shadmi}},\
  }\bibfield  {title} {\enquote {\bibinfo {title} {Testing gluon
  self-interactions in three-jet events at hadron colliders},}\ }\href
  {\doibase 10.1016/0550-3213(94)90563-0} {\bibfield  {journal} {\bibinfo
  {journal} {Nuclear Physics B}\ }\textbf {\bibinfo {volume} {423}},\ \bibinfo
  {pages} {3–32} (\bibinfo {year} {1994})}\BibitemShut {NoStop}%
\bibitem [{\citenamefont {Degrande}\ and\ \citenamefont
  {Maltoni}(2021)}]{Degrande:2021rev}%
  \BibitemOpen
  \bibfield  {author} {\bibinfo {author} {\bibfnamefont {C.}~\bibnamefont
  {Degrande}}\ and\ \bibinfo {author} {\bibfnamefont {M.}~\bibnamefont
  {Maltoni}},\ }\bibfield  {title} {\enquote {\bibinfo {title} {{Reviving the
  interference: Framework and proof-of-principle for the anomalous gluon
  self-interaction in the SMEFT}},}\ }\href {\doibase
  10.1103/physrevd.103.095009} {\bibfield  {journal} {\bibinfo  {journal}
  {Physical Review D}\ }\textbf {\bibinfo {volume} {103}} (\bibinfo {year}
  {2021}),\ 10.1103/physrevd.103.095009}\BibitemShut {NoStop}%
\bibitem [{\citenamefont {Degrande}\ \emph {et~al.}(2021)\citenamefont
  {Degrande}, \citenamefont {Durieux}, \citenamefont {Maltoni}, \citenamefont
  {Mimasu}, \citenamefont {Vryonidou},\ and\ \citenamefont
  {Zhang}}]{Degrande:2021autom}%
  \BibitemOpen
  \bibfield  {author} {\bibinfo {author} {\bibfnamefont {C.}~\bibnamefont
  {Degrande}}, \bibinfo {author} {\bibfnamefont {G.}~\bibnamefont {Durieux}},
  \bibinfo {author} {\bibfnamefont {F.}~\bibnamefont {Maltoni}}, \bibinfo
  {author} {\bibfnamefont {K.}~\bibnamefont {Mimasu}}, \bibinfo {author}
  {\bibfnamefont {E.}~\bibnamefont {Vryonidou}}, \ and\ \bibinfo {author}
  {\bibfnamefont {C.}~\bibnamefont {Zhang}},\ }\bibfield  {title} {\enquote
  {\bibinfo {title} {Automated one-loop computations in the standard model
  effective field theory},}\ }\href {\doibase 10.1103/physrevd.103.096024}
  {\bibfield  {journal} {\bibinfo  {journal} {Physical Review D}\ }\textbf
  {\bibinfo {volume} {103}} (\bibinfo {year} {2021}),\
  10.1103/physrevd.103.096024}\BibitemShut {NoStop}%
\bibitem [{\citenamefont {Bhardwaj}\ \emph {et~al.}(2022)\citenamefont
  {Bhardwaj}, \citenamefont {Englert}, \citenamefont {Hankache},\ and\
  \citenamefont {Pilkington}}]{CPasymmML:2022higgs}%
  \BibitemOpen
  \bibfield  {author} {\bibinfo {author} {\bibfnamefont {A.}~\bibnamefont
  {Bhardwaj}}, \bibinfo {author} {\bibfnamefont {C.}~\bibnamefont {Englert}},
  \bibinfo {author} {\bibfnamefont {R.}~\bibnamefont {Hankache}}, \ and\
  \bibinfo {author} {\bibfnamefont {A.D.}\ \bibnamefont {Pilkington}},\
  }\bibfield  {title} {\enquote {\bibinfo {title} {{Machine-enhanced
  CP-asymmetries in the Higgs sector}},}\ }\href {\doibase
  10.1016/j.physletb.2022.137246} {\bibfield  {journal} {\bibinfo  {journal}
  {Physics Letters B}\ }\textbf {\bibinfo {volume} {832}} (\bibinfo {year}
  {2022}),\ 10.1016/j.physletb.2022.137246}\BibitemShut {NoStop}%
\bibitem [{\citenamefont {Clarke~Hall}\ \emph {et~al.}(2023)\citenamefont
  {Clarke~Hall}, \citenamefont {Criddle}, \citenamefont {Crossland},
  \citenamefont {Englert}, \citenamefont {Forbes}, \citenamefont {Hankache},\
  and\ \citenamefont {Pilkington}}]{CPasymmML:2022ew}%
  \BibitemOpen
  \bibfield  {author} {\bibinfo {author} {\bibfnamefont {N.}~\bibnamefont
  {Clarke~Hall}}, \bibinfo {author} {\bibfnamefont {I.}~\bibnamefont
  {Criddle}}, \bibinfo {author} {\bibfnamefont {A}~\bibnamefont {Crossland}},
  \bibinfo {author} {\bibfnamefont {C.}~\bibnamefont {Englert}}, \bibinfo
  {author} {\bibfnamefont {P.}~\bibnamefont {Forbes}}, \bibinfo {author}
  {\bibfnamefont {R.}~\bibnamefont {Hankache}}, \ and\ \bibinfo {author}
  {\bibfnamefont {A.D.}\ \bibnamefont {Pilkington}},\ }\bibfield  {title}
  {\enquote {\bibinfo {title} {{Machine-enhanced CP-asymmetries in the
  electroweak sector}},}\ }\href@noop {} {\  (\bibinfo {year} {2023})},\
  \Eprint {http://arxiv.org/abs/2209.05143} {arXiv:2209.05143 [hep-ph]}
  \BibitemShut {NoStop}%
\bibitem [{\citenamefont {Alwall}\ \emph {et~al.}(2014)\citenamefont {Alwall},
  \citenamefont {Frederix}, \citenamefont {Frixione}, \citenamefont {Hirschi},
  \citenamefont {Maltoni}, \citenamefont {Mattelaer}, \citenamefont {Shao},
  \citenamefont {Stelzer}, \citenamefont {Torrielli},\ and\ \citenamefont
  {Zaro}}]{Alwall:2014}%
  \BibitemOpen
  \bibfield  {author} {\bibinfo {author} {\bibfnamefont {J.}~\bibnamefont
  {Alwall}}, \bibinfo {author} {\bibfnamefont {R.}~\bibnamefont {Frederix}},
  \bibinfo {author} {\bibfnamefont {S.}~\bibnamefont {Frixione}}, \bibinfo
  {author} {\bibfnamefont {V.}~\bibnamefont {Hirschi}}, \bibinfo {author}
  {\bibfnamefont {F.}~\bibnamefont {Maltoni}}, \bibinfo {author} {\bibfnamefont
  {O.}~\bibnamefont {Mattelaer}}, \bibinfo {author} {\bibfnamefont {H.-S.}\
  \bibnamefont {Shao}}, \bibinfo {author} {\bibfnamefont {T.}~\bibnamefont
  {Stelzer}}, \bibinfo {author} {\bibfnamefont {P.}~\bibnamefont {Torrielli}},
  \ and\ \bibinfo {author} {\bibfnamefont {M.}~\bibnamefont {Zaro}},\
  }\bibfield  {title} {\enquote {\bibinfo {title} {The automated computation of
  tree-level and next-to-leading order differential cross sections, and their
  matching to parton shower simulations},}\ }\href {\doibase
  10.1007/jhep07(2014)079} {\bibfield  {journal} {\bibinfo  {journal} {Journal
  of High Energy Physics}\ }\textbf {\bibinfo {volume} {2014}} (\bibinfo {year}
  {2014}),\ 10.1007/jhep07(2014)079}\BibitemShut {NoStop}%
\bibitem [{\citenamefont {Degrande}\ \emph {et~al.}(2012)\citenamefont
  {Degrande}, \citenamefont {Duhr}, \citenamefont {Fuks}, \citenamefont
  {Grellscheid}, \citenamefont {Mattelaer},\ and\ \citenamefont
  {Reiter}}]{Degrande:2011ua}%
  \BibitemOpen
  \bibfield  {author} {\bibinfo {author} {\bibfnamefont {C.}~\bibnamefont
  {Degrande}}, \bibinfo {author} {\bibfnamefont {C.}~\bibnamefont {Duhr}},
  \bibinfo {author} {\bibfnamefont {B.}~\bibnamefont {Fuks}}, \bibinfo {author}
  {\bibfnamefont {D.}~\bibnamefont {Grellscheid}}, \bibinfo {author}
  {\bibfnamefont {O.}~\bibnamefont {Mattelaer}}, \ and\ \bibinfo {author}
  {\bibfnamefont {T.}~\bibnamefont {Reiter}},\ }\bibfield  {title} {\enquote
  {\bibinfo {title} {{UFO - The Universal FeynRules Output}},}\ }\href
  {\doibase 10.1016/j.cpc.2012.01.022} {\bibfield  {journal} {\bibinfo
  {journal} {Comput. Phys. Commun.}\ }\textbf {\bibinfo {volume} {183}},\
  \bibinfo {pages} {1201--1214} (\bibinfo {year} {2012})},\ \Eprint
  {http://arxiv.org/abs/1108.2040} {arXiv:1108.2040 [hep-ph]} \BibitemShut
  {NoStop}%
\bibitem [{\citenamefont {Alloul}\ \emph {et~al.}(2014)\citenamefont {Alloul},
  \citenamefont {Christensen}, \citenamefont {Degrande}, \citenamefont {Duhr},\
  and\ \citenamefont {Fuks}}]{Alloul:2013bka}%
  \BibitemOpen
  \bibfield  {author} {\bibinfo {author} {\bibfnamefont {A.}~\bibnamefont
  {Alloul}}, \bibinfo {author} {\bibfnamefont {N.D.}\ \bibnamefont
  {Christensen}}, \bibinfo {author} {\bibfnamefont {C.}~\bibnamefont
  {Degrande}}, \bibinfo {author} {\bibfnamefont {C.}~\bibnamefont {Duhr}}, \
  and\ \bibinfo {author} {\bibfnamefont {B.}~\bibnamefont {Fuks}},\ }\bibfield
  {title} {\enquote {\bibinfo {title} {{FeynRules 2.0 - A complete toolbox for
  tree-level phenomenology}},}\ }\href {\doibase 10.1016/j.cpc.2014.04.012}
  {\bibfield  {journal} {\bibinfo  {journal} {Comput. Phys. Commun.}\ }\textbf
  {\bibinfo {volume} {185}},\ \bibinfo {pages} {2250--2300} (\bibinfo {year}
  {2014})},\ \Eprint {http://arxiv.org/abs/1310.1921} {arXiv:1310.1921
  [hep-ph]} \BibitemShut {NoStop}%
\bibitem [{\citenamefont {Degrande}(2015)}]{Degrande:2015nloct}%
  \BibitemOpen
  \bibfield  {author} {\bibinfo {author} {\bibfnamefont {C.}~\bibnamefont
  {Degrande}},\ }\bibfield  {title} {\enquote {\bibinfo {title} {{Automatic
  evaluation of UV and R2 terms for beyond the Standard Model Lagrangians: a
  proof-of-principle}},}\ }\href {\doibase 10.1016/j.cpc.2015.08.015}
  {\bibfield  {journal} {\bibinfo  {journal} {Computer Physics Communications}\
  }\textbf {\bibinfo {volume} {197}},\ \bibinfo {pages} {239--262} (\bibinfo
  {year} {2015})},\ \Eprint {http://arxiv.org/abs/1406.3030} {arXiv:1406.3030
  [hep-ph]} \BibitemShut {NoStop}%
\bibitem [{\citenamefont {Ball}\ \emph {et~al.}(2015)\citenamefont {Ball},
  \citenamefont {Bertone}, \citenamefont {Carrazza}, \citenamefont {Deans},
  \citenamefont {Del~Debbio}, \citenamefont {Forte}, \citenamefont {Guffanti},
  \citenamefont {Hartland}, \citenamefont {Latorre},\ and\ \citenamefont
  {et~al.}}]{Ball:2015}%
  \BibitemOpen
  \bibfield  {author} {\bibinfo {author} {\bibfnamefont {R.D.}\ \bibnamefont
  {Ball}}, \bibinfo {author} {\bibfnamefont {V.}~\bibnamefont {Bertone}},
  \bibinfo {author} {\bibfnamefont {S.}~\bibnamefont {Carrazza}}, \bibinfo
  {author} {\bibfnamefont {C.S.}\ \bibnamefont {Deans}}, \bibinfo {author}
  {\bibfnamefont {L.}~\bibnamefont {Del~Debbio}}, \bibinfo {author}
  {\bibfnamefont {S.}~\bibnamefont {Forte}}, \bibinfo {author} {\bibfnamefont
  {A.}~\bibnamefont {Guffanti}}, \bibinfo {author} {\bibfnamefont {N.P.}\
  \bibnamefont {Hartland}}, \bibinfo {author} {\bibfnamefont {J.I.}\
  \bibnamefont {Latorre}}, \ and\ \bibinfo {author} {\bibnamefont {et~al.}},\
  }\bibfield  {title} {\enquote {\bibinfo {title} {{Parton distributions for
  the LHC run II}},}\ }\href {\doibase 10.1007/jhep04(2015)040} {\bibfield
  {journal} {\bibinfo  {journal} {Journal of High Energy Physics}\ }\textbf
  {\bibinfo {volume} {2015}} (\bibinfo {year} {2015}),\
  10.1007/jhep04(2015)040}\BibitemShut {NoStop}%
\bibitem [{\citenamefont {Frixione}\ and\ \citenamefont
  {Webber}(2002)}]{Frixione:2002}%
  \BibitemOpen
  \bibfield  {author} {\bibinfo {author} {\bibfnamefont {S.}~\bibnamefont
  {Frixione}}\ and\ \bibinfo {author} {\bibfnamefont {B.R.}\ \bibnamefont
  {Webber}},\ }\bibfield  {title} {\enquote {\bibinfo {title} {{Matching NLO
  QCD computations and parton shower simulations}},}\ }\href {\doibase
  10.1088/1126-6708/2002/06/029} {\bibfield  {journal} {\bibinfo  {journal}
  {Journal of High Energy Physics}\ }\textbf {\bibinfo {volume} {2002}},\
  \bibinfo {pages} {029–029} (\bibinfo {year} {2002})}\BibitemShut {NoStop}%
\bibitem [{\citenamefont {Sjöstrand}\ \emph {et~al.}(2015)\citenamefont
  {Sjöstrand}, \citenamefont {Ask}, \citenamefont {Christiansen},
  \citenamefont {Corke}, \citenamefont {Desai}, \citenamefont {Ilten},
  \citenamefont {Mrenna}, \citenamefont {Prestel}, \citenamefont {Rasmussen},\
  and\ \citenamefont {Skands}}]{Pythia:2015}%
  \BibitemOpen
  \bibfield  {author} {\bibinfo {author} {\bibfnamefont {T.}~\bibnamefont
  {Sjöstrand}}, \bibinfo {author} {\bibfnamefont {S.}~\bibnamefont {Ask}},
  \bibinfo {author} {\bibfnamefont {J.R.}\ \bibnamefont {Christiansen}},
  \bibinfo {author} {\bibfnamefont {R.}~\bibnamefont {Corke}}, \bibinfo
  {author} {\bibfnamefont {N.}~\bibnamefont {Desai}}, \bibinfo {author}
  {\bibfnamefont {P.}~\bibnamefont {Ilten}}, \bibinfo {author} {\bibfnamefont
  {S.}~\bibnamefont {Mrenna}}, \bibinfo {author} {\bibfnamefont
  {S.}~\bibnamefont {Prestel}}, \bibinfo {author} {\bibfnamefont {C.O.}\
  \bibnamefont {Rasmussen}}, \ and\ \bibinfo {author} {\bibfnamefont {P.Z.}\
  \bibnamefont {Skands}},\ }\bibfield  {title} {\enquote {\bibinfo {title} {{An
  introduction to PYTHIA 8.2}},}\ }\href {\doibase 10.1016/j.cpc.2015.01.024}
  {\bibfield  {journal} {\bibinfo  {journal} {Computer Physics Communications}\
  }\textbf {\bibinfo {volume} {191}},\ \bibinfo {pages} {159–177} (\bibinfo
  {year} {2015})}\BibitemShut {NoStop}%
\bibitem [{\citenamefont {Bähr}\ \emph {et~al.}(2008)\citenamefont {Bähr},
  \citenamefont {Gieseke}, \citenamefont {Gigg}, \citenamefont {Grellscheid},
  \citenamefont {Hamilton}, \citenamefont {Latunde-Dada}, \citenamefont
  {Plätzer}, \citenamefont {Richardson}, \citenamefont {Seymour},
  \citenamefont {Sherstnev},\ and\ \citenamefont {Webber}}]{Herwig:2008}%
  \BibitemOpen
  \bibfield  {author} {\bibinfo {author} {\bibfnamefont {M.}~\bibnamefont
  {Bähr}}, \bibinfo {author} {\bibfnamefont {S.}~\bibnamefont {Gieseke}},
  \bibinfo {author} {\bibfnamefont {M.A.}\ \bibnamefont {Gigg}}, \bibinfo
  {author} {\bibfnamefont {D.}~\bibnamefont {Grellscheid}}, \bibinfo {author}
  {\bibfnamefont {K.}~\bibnamefont {Hamilton}}, \bibinfo {author}
  {\bibfnamefont {O.}~\bibnamefont {Latunde-Dada}}, \bibinfo {author}
  {\bibfnamefont {S.}~\bibnamefont {Plätzer}}, \bibinfo {author}
  {\bibfnamefont {P.}~\bibnamefont {Richardson}}, \bibinfo {author}
  {\bibfnamefont {M.H.}\ \bibnamefont {Seymour}}, \bibinfo {author}
  {\bibfnamefont {A.}~\bibnamefont {Sherstnev}}, \ and\ \bibinfo {author}
  {\bibfnamefont {B.R.}\ \bibnamefont {Webber}},\ }\bibfield  {title} {\enquote
  {\bibinfo {title} {{Herwig++ physics and manual}},}\ }\href {\doibase
  10.1140/epjc/s10052-008-0798-9} {\bibfield  {journal} {\bibinfo  {journal}
  {The European Physical Journal C}\ }\textbf {\bibinfo {volume} {58}},\
  \bibinfo {pages} {639–707} (\bibinfo {year} {2008})}\BibitemShut {NoStop}%
\bibitem [{\citenamefont {Bellm}\ \emph {et~al.}(2016)\citenamefont {Bellm},
  \citenamefont {Gieseke}, \citenamefont {Grellscheid}, \citenamefont
  {Plätzer}, \citenamefont {Rauch}, \citenamefont {Reuschle}, \citenamefont
  {Richardson}, \citenamefont {Schichtel}, \citenamefont {Seymour},
  \citenamefont {Siódmok}, \citenamefont {Wilcock}, \citenamefont {Fischer},
  \citenamefont {Harrendorf}, \citenamefont {Nail}, \citenamefont
  {Papaefstathiou},\ and\ \citenamefont {Rauch}}]{Herwig:2016}%
  \BibitemOpen
  \bibfield  {author} {\bibinfo {author} {\bibfnamefont {J.}~\bibnamefont
  {Bellm}}, \bibinfo {author} {\bibfnamefont {S.}~\bibnamefont {Gieseke}},
  \bibinfo {author} {\bibfnamefont {D.}~\bibnamefont {Grellscheid}}, \bibinfo
  {author} {\bibfnamefont {S.}~\bibnamefont {Plätzer}}, \bibinfo {author}
  {\bibfnamefont {M.}~\bibnamefont {Rauch}}, \bibinfo {author} {\bibfnamefont
  {C.}~\bibnamefont {Reuschle}}, \bibinfo {author} {\bibfnamefont
  {P.}~\bibnamefont {Richardson}}, \bibinfo {author} {\bibfnamefont
  {P.}~\bibnamefont {Schichtel}}, \bibinfo {author} {\bibfnamefont {M.H.}\
  \bibnamefont {Seymour}}, \bibinfo {author} {\bibfnamefont {A.}~\bibnamefont
  {Siódmok}}, \bibinfo {author} {\bibfnamefont {A.}~\bibnamefont {Wilcock}},
  \bibinfo {author} {\bibfnamefont {N.}~\bibnamefont {Fischer}}, \bibinfo
  {author} {\bibfnamefont {M.A.}\ \bibnamefont {Harrendorf}}, \bibinfo {author}
  {\bibfnamefont {G.}~\bibnamefont {Nail}}, \bibinfo {author} {\bibfnamefont
  {A.}~\bibnamefont {Papaefstathiou}}, \ and\ \bibinfo {author} {\bibfnamefont
  {D.}~\bibnamefont {Rauch}},\ }\bibfield  {title} {\enquote {\bibinfo {title}
  {{Herwig 7.0/Herwig++ 3.0 release note}},}\ }\href {\doibase
  10.1140/epjc/s10052-016-4018-8} {\bibfield  {journal} {\bibinfo  {journal}
  {The European Physical Journal C}\ }\textbf {\bibinfo {volume} {76}}
  (\bibinfo {year} {2016}),\ 10.1140/epjc/s10052-016-4018-8}\BibitemShut
  {NoStop}%
\bibitem [{\citenamefont {Cacciari}\ \emph {et~al.}(2012)\citenamefont
  {Cacciari}, \citenamefont {Salam},\ and\ \citenamefont
  {Soyez}}]{Fastjet:2012}%
  \BibitemOpen
  \bibfield  {author} {\bibinfo {author} {\bibfnamefont {M.}~\bibnamefont
  {Cacciari}}, \bibinfo {author} {\bibfnamefont {G.P.}\ \bibnamefont {Salam}},
  \ and\ \bibinfo {author} {\bibfnamefont {G.}~\bibnamefont {Soyez}},\
  }\bibfield  {title} {\enquote {\bibinfo {title} {{FastJet user manual (for
  version 3.0.2)}},}\ }\href {\doibase 10.1140/epjc/s10052-012-1896-2}
  {\bibfield  {journal} {\bibinfo  {journal} {The European Physical Journal C}\
  }\textbf {\bibinfo {volume} {72}} (\bibinfo {year} {2012}),\
  10.1140/epjc/s10052-012-1896-2}\BibitemShut {NoStop}%
\bibitem [{\citenamefont {Catani}\ \emph {et~al.}(1993)\citenamefont {Catani},
  \citenamefont {Dokshitzer}, \citenamefont {Seymour},\ and\ \citenamefont
  {Webber}}]{kt_1}%
  \BibitemOpen
  \bibfield  {author} {\bibinfo {author} {\bibfnamefont {S.}~\bibnamefont
  {Catani}}, \bibinfo {author} {\bibfnamefont {Yuri~L.}\ \bibnamefont
  {Dokshitzer}}, \bibinfo {author} {\bibfnamefont {M.~H.}\ \bibnamefont
  {Seymour}}, \ and\ \bibinfo {author} {\bibfnamefont {B.~R.}\ \bibnamefont
  {Webber}},\ }\bibfield  {title} {\enquote {\bibinfo {title} {{Longitudinally
  invariant $K_t$ clustering algorithms for hadron hadron collisions}},}\
  }\href {\doibase 10.1016/0550-3213(93)90166-M} {\bibfield  {journal}
  {\bibinfo  {journal} {Nucl. Phys. B}\ }\textbf {\bibinfo {volume} {406}},\
  \bibinfo {pages} {187--224} (\bibinfo {year} {1993})}\BibitemShut {NoStop}%
\bibitem [{\citenamefont {Ellis}\ and\ \citenamefont {Soper}(1993)}]{kt_2}%
  \BibitemOpen
  \bibfield  {author} {\bibinfo {author} {\bibfnamefont {S.D.}\ \bibnamefont
  {Ellis}}\ and\ \bibinfo {author} {\bibfnamefont {D.E.}\ \bibnamefont
  {Soper}},\ }\bibfield  {title} {\enquote {\bibinfo {title} {{Successive
  combination jet algorithm for hadron collisions}},}\ }\href {\doibase
  10.1103/physrevd.48.3160} {\bibfield  {journal} {\bibinfo  {journal}
  {Physical Review D}\ }\textbf {\bibinfo {volume} {48}},\ \bibinfo {pages}
  {3160–3166} (\bibinfo {year} {1993})}\BibitemShut {NoStop}%
\bibitem [{\citenamefont {Cacciari}\ \emph {et~al.}(2008)\citenamefont
  {Cacciari}, \citenamefont {Salam},\ and\ \citenamefont
  {Soyez}}]{AntiKt:2008}%
  \BibitemOpen
  \bibfield  {author} {\bibinfo {author} {\bibfnamefont {M.}~\bibnamefont
  {Cacciari}}, \bibinfo {author} {\bibfnamefont {G.P.}\ \bibnamefont {Salam}},
  \ and\ \bibinfo {author} {\bibfnamefont {G.}~\bibnamefont {Soyez}},\
  }\bibfield  {title} {\enquote {\bibinfo {title} {{The anti-kt jet clustering
  algorithm}},}\ }\href {\doibase 10.1088/1126-6708/2008/04/063} {\bibfield
  {journal} {\bibinfo  {journal} {Journal of High Energy Physics}\ }\textbf
  {\bibinfo {volume} {2008}},\ \bibinfo {pages} {063–063} (\bibinfo {year}
  {2008})}\BibitemShut {NoStop}%
\bibitem [{\citenamefont {Frederix}\ \emph {et~al.}(2012)\citenamefont
  {Frederix}, \citenamefont {Frixione}, \citenamefont {Hirschi}, \citenamefont
  {Maltoni}, \citenamefont {Pittau},\ and\ \citenamefont
  {Torrielli}}]{Frederix:2012}%
  \BibitemOpen
  \bibfield  {author} {\bibinfo {author} {\bibfnamefont {R.}~\bibnamefont
  {Frederix}}, \bibinfo {author} {\bibfnamefont {S.}~\bibnamefont {Frixione}},
  \bibinfo {author} {\bibfnamefont {V.}~\bibnamefont {Hirschi}}, \bibinfo
  {author} {\bibfnamefont {F.}~\bibnamefont {Maltoni}}, \bibinfo {author}
  {\bibfnamefont {R.}~\bibnamefont {Pittau}}, \ and\ \bibinfo {author}
  {\bibfnamefont {P.}~\bibnamefont {Torrielli}},\ }\bibfield  {title} {\enquote
  {\bibinfo {title} {{Four-lepton production at hadron colliders: aMC@NLO
  predictions with theoretical uncertainties}},}\ }\href {\doibase
  10.1007/jhep02(2012)099} {\bibfield  {journal} {\bibinfo  {journal} {Journal
  of High Energy Physics}\ }\textbf {\bibinfo {volume} {2012}} (\bibinfo {year}
  {2012}),\ 10.1007/jhep02(2012)099}\BibitemShut {NoStop}%
\bibitem [{\citenamefont {Li}\ \emph {et~al.}(2021)\citenamefont {Li},
  \citenamefont {Ren}, \citenamefont {Shu}, \citenamefont {Xiao}, \citenamefont
  {Yu},\ and\ \citenamefont {Zheng}}]{dim8_basis}%
  \BibitemOpen
  \bibfield  {author} {\bibinfo {author} {\bibfnamefont {H.-L.}\ \bibnamefont
  {Li}}, \bibinfo {author} {\bibfnamefont {Z.}~\bibnamefont {Ren}}, \bibinfo
  {author} {\bibfnamefont {J.}~\bibnamefont {Shu}}, \bibinfo {author}
  {\bibfnamefont {M.-L.}\ \bibnamefont {Xiao}}, \bibinfo {author}
  {\bibfnamefont {J.-H.}\ \bibnamefont {Yu}}, \ and\ \bibinfo {author}
  {\bibfnamefont {Y.-H.}\ \bibnamefont {Zheng}},\ }\bibfield  {title} {\enquote
  {\bibinfo {title} {{Complete set of dimension-eight operators in the standard
  model effective field theory}},}\ }\href {\doibase
  10.1103/physrevd.104.015026} {\bibfield  {journal} {\bibinfo  {journal}
  {Physical Review D}\ }\textbf {\bibinfo {volume} {104}} (\bibinfo {year}
  {2021}),\ 10.1103/physrevd.104.015026}\BibitemShut {NoStop}%
\bibitem [{\citenamefont {Collaboration}(2021{\natexlab{a}})}]{Atlas:2020zjj}%
  \BibitemOpen
  \bibfield  {author} {\bibinfo {author} {\bibfnamefont {ATLAS}\ \bibnamefont
  {Collaboration}},\ }\bibfield  {title} {\enquote {\bibinfo {title}
  {{Differential cross-section measurements for the electroweak production of
  dijets in association with a Z boson in proton–proton collisions at
  ATLAS}},}\ }\href {\doibase 10.1140/epjc/s10052-020-08734-w} {\bibfield
  {journal} {\bibinfo  {journal} {The European Physical Journal C}\ }\textbf
  {\bibinfo {volume} {81}} (\bibinfo {year} {2021}{\natexlab{a}}),\
  10.1140/epjc/s10052-020-08734-w}\BibitemShut {NoStop}%
\bibitem [{\citenamefont
  {Collaboration}(2021{\natexlab{b}})}]{ATL-PHYS-PUB-2021-022}%
  \BibitemOpen
  \bibfield  {author} {\bibinfo {author} {\bibfnamefont {ATLAS}\ \bibnamefont
  {Collaboration}},\ }\bibfield  {title} {\enquote {\bibinfo {title} {{Combined
  effective field theory interpretation of differential cross-sections
  measurements of $WW$, $WZ$, 4$\ell$, and $Z$-plus-two-jets production using
  ATLAS data}},}\ }\href {https://cds.cern.ch/record/2776648} {\bibfield
  {journal} {\bibinfo  {journal} {{ATL-PHYS-PUB-2021-022}}\ } (\bibinfo {year}
  {2021}{\natexlab{b}})}\BibitemShut {NoStop}%
\bibitem [{\citenamefont {Collaboration}(2019{\natexlab{a}})}]{WWjj_2019}%
  \BibitemOpen
  \bibfield  {author} {\bibinfo {author} {\bibfnamefont {ATLAS}\ \bibnamefont
  {Collaboration}},\ }\bibfield  {title} {\enquote {\bibinfo {title}
  {{Observation of Electroweak Production of a Same-Sign $W$ boson pair in
  association with two jets in $pp$ collisions at $\sqrt{s}=13$ TeV with the
  ATLAS detector}},}\ }\href {\doibase 10.1103/physrevlett.123.161801}
  {\bibfield  {journal} {\bibinfo  {journal} {Physical Review Letters}\
  }\textbf {\bibinfo {volume} {123}} (\bibinfo {year} {2019}{\natexlab{a}}),\
  10.1103/physrevlett.123.161801}\BibitemShut {NoStop}%
\bibitem [{\citenamefont {Collaboration}(2019{\natexlab{b}})}]{WWjj_1_2019}%
  \BibitemOpen
  \bibfield  {author} {\bibinfo {author} {\bibfnamefont {ATLAS}\ \bibnamefont
  {Collaboration}},\ }\bibfield  {title} {\enquote {\bibinfo {title}
  {{Modelling of the vector boson scattering process $pp\rightarrow W^\pm W^\pm
  jj$ in Monte Carlo generators in ATLAS}},}\ }\href
  {https://inspirehep.net/literature/1795235} {\bibfield  {journal} {\bibinfo
  {journal} {{ATL-PHYS-PUB-2019-004}}\ } (\bibinfo {year}
  {2019}{\natexlab{b}})}\BibitemShut {NoStop}%
\bibitem [{\citenamefont {Höche}\ \emph {et~al.}(2021)\citenamefont {Höche},
  \citenamefont {Mrenna}, \citenamefont {Payne}, \citenamefont {Preuss},\ and\
  \citenamefont {Skands}}]{Hoche:2021}%
  \BibitemOpen
  \bibfield  {author} {\bibinfo {author} {\bibfnamefont {S.}~\bibnamefont
  {Höche}}, \bibinfo {author} {\bibfnamefont {S.}~\bibnamefont {Mrenna}},
  \bibinfo {author} {\bibfnamefont {S.}~\bibnamefont {Payne}}, \bibinfo
  {author} {\bibfnamefont {C.T.}\ \bibnamefont {Preuss}}, \ and\ \bibinfo
  {author} {\bibfnamefont {P.}~\bibnamefont {Skands}},\ }\bibfield  {title}
  {\enquote {\bibinfo {title} {{A Study of QCD Radiation in VBF Higgs
  Production with Vincia and Pythia}},}\ }\href@noop {} {\  (\bibinfo {year}
  {2021})},\ \Eprint {http://arxiv.org/abs/2106.10987} {arXiv:2106.10987
  [hep-ph]} \BibitemShut {NoStop}%
\bibitem [{\citenamefont {Jäger}\ \emph {et~al.}(2020)\citenamefont {Jäger},
  \citenamefont {Karlberg}, \citenamefont {Plätzer}, \citenamefont
  {Scheller},\ and\ \citenamefont {Zaro}}]{Jager:2020}%
  \BibitemOpen
  \bibfield  {author} {\bibinfo {author} {\bibfnamefont {B.}~\bibnamefont
  {Jäger}}, \bibinfo {author} {\bibfnamefont {A.}~\bibnamefont {Karlberg}},
  \bibinfo {author} {\bibfnamefont {S.}~\bibnamefont {Plätzer}}, \bibinfo
  {author} {\bibfnamefont {J.}~\bibnamefont {Scheller}}, \ and\ \bibinfo
  {author} {\bibfnamefont {M.}~\bibnamefont {Zaro}},\ }\bibfield  {title}
  {\enquote {\bibinfo {title} {{Parton-shower effects in Higgs production via
  vector-boson fusion}},}\ }\href {\doibase 10.1140/epjc/s10052-020-8326-7}
  {\bibfield  {journal} {\bibinfo  {journal} {The European Physical Journal C}\
  }\textbf {\bibinfo {volume} {80}} (\bibinfo {year} {2020}),\
  10.1140/epjc/s10052-020-8326-7}\BibitemShut {NoStop}%
\bibitem [{\citenamefont {Baglio}\ \emph {et~al.}(2011)\citenamefont {Baglio},
  \citenamefont {Bellm}, \citenamefont {Bozzi}, \citenamefont {Brieg},
  \citenamefont {Campanario}, \citenamefont {Englert}, \citenamefont {Feigl},
  \citenamefont {Frank}, \citenamefont {Figy}, \citenamefont {Geyer},
  \citenamefont {Hackstein}, \citenamefont {Hankele}, \citenamefont {Jäger},
  \citenamefont {Kerner}, \citenamefont {Kubocz}, \citenamefont {Ninh},
  \citenamefont {Oleari}, \citenamefont {Palmer}, \citenamefont {Plätzer},
  \citenamefont {Rauch}, \citenamefont {Roth}, \citenamefont {Rzehak},
  \citenamefont {Schissler}, \citenamefont {Schlimpert}, \citenamefont
  {Spannowsky}, \citenamefont {Worek},\ and\ \citenamefont
  {Zeppenfeld}}]{Vbfnlo}%
  \BibitemOpen
  \bibfield  {author} {\bibinfo {author} {\bibfnamefont {J.}~\bibnamefont
  {Baglio}}, \bibinfo {author} {\bibfnamefont {J.}~\bibnamefont {Bellm}},
  \bibinfo {author} {\bibfnamefont {G.}~\bibnamefont {Bozzi}}, \bibinfo
  {author} {\bibfnamefont {M.}~\bibnamefont {Brieg}}, \bibinfo {author}
  {\bibfnamefont {F.}~\bibnamefont {Campanario}}, \bibinfo {author}
  {\bibfnamefont {C.}~\bibnamefont {Englert}}, \bibinfo {author} {\bibfnamefont
  {B.}~\bibnamefont {Feigl}}, \bibinfo {author} {\bibfnamefont
  {J.}~\bibnamefont {Frank}}, \bibinfo {author} {\bibfnamefont
  {T.}~\bibnamefont {Figy}}, \bibinfo {author} {\bibfnamefont {F.}~\bibnamefont
  {Geyer}}, \bibinfo {author} {\bibfnamefont {C.}~\bibnamefont {Hackstein}},
  \bibinfo {author} {\bibfnamefont {V.}~\bibnamefont {Hankele}}, \bibinfo
  {author} {\bibfnamefont {B.}~\bibnamefont {Jäger}}, \bibinfo {author}
  {\bibfnamefont {M.}~\bibnamefont {Kerner}}, \bibinfo {author} {\bibfnamefont
  {M.}~\bibnamefont {Kubocz}}, \bibinfo {author} {\bibfnamefont {L.D.}\
  \bibnamefont {Ninh}}, \bibinfo {author} {\bibfnamefont {C.}~\bibnamefont
  {Oleari}}, \bibinfo {author} {\bibfnamefont {S.}~\bibnamefont {Palmer}},
  \bibinfo {author} {\bibfnamefont {S.}~\bibnamefont {Plätzer}}, \bibinfo
  {author} {\bibfnamefont {M.}~\bibnamefont {Rauch}}, \bibinfo {author}
  {\bibfnamefont {R.}~\bibnamefont {Roth}}, \bibinfo {author} {\bibfnamefont
  {H.}~\bibnamefont {Rzehak}}, \bibinfo {author} {\bibfnamefont
  {F.}~\bibnamefont {Schissler}}, \bibinfo {author} {\bibfnamefont
  {O.}~\bibnamefont {Schlimpert}}, \bibinfo {author} {\bibfnamefont
  {M.}~\bibnamefont {Spannowsky}}, \bibinfo {author} {\bibfnamefont
  {M.}~\bibnamefont {Worek}}, \ and\ \bibinfo {author} {\bibfnamefont
  {D.}~\bibnamefont {Zeppenfeld}},\ }\bibfield  {title} {\enquote {\bibinfo
  {title} {{VBFNLO: A parton level Monte Carlo for processes with electroweak
  bosons -- Manual for Version 2.7.0}},}\ }\href {\doibase
  10.48550/arXiv.1107.4038} {\  (\bibinfo {year} {2011}),\
  10.48550/arXiv.1107.4038},\ \Eprint {http://arxiv.org/abs/1107.4038}
  {arXiv:1107.4038} \BibitemShut {NoStop}%
\bibitem [{\citenamefont {Aoude}\ and\ \citenamefont
  {Shepherd}(2019)}]{rafa_will_hadr}%
  \BibitemOpen
  \bibfield  {author} {\bibinfo {author} {\bibfnamefont {R.}~\bibnamefont
  {Aoude}}\ and\ \bibinfo {author} {\bibfnamefont {W.}~\bibnamefont
  {Shepherd}},\ }\bibfield  {title} {\enquote {\bibinfo {title} {{Jet
  substructure measurements of interference in non-interfering SMEFT
  effects}},}\ }\href {\doibase 10.1007/jhep08(2019)009} {\bibfield  {journal}
  {\bibinfo  {journal} {Journal of High Energy Physics}\ }\textbf {\bibinfo
  {volume} {2019}} (\bibinfo {year} {2019}),\
  10.1007/jhep08(2019)009}\BibitemShut {NoStop}%
\bibitem [{\citenamefont {Aaboud}\ \emph {et~al.}(2019)\citenamefont {Aaboud},
  \citenamefont {Aad}, \citenamefont {Abbott}, \citenamefont {Abdinov},
  \citenamefont {Abeloos}, \citenamefont {Abhayasinghe}, \citenamefont {Abidi},
  \citenamefont {AbouZeid}, \citenamefont {Abraham},\ and\ \citenamefont
  {et~al.}}]{Atlas:2019wz}%
  \BibitemOpen
  \bibfield  {author} {\bibinfo {author} {\bibfnamefont {M.}~\bibnamefont
  {Aaboud}}, \bibinfo {author} {\bibfnamefont {G.}~\bibnamefont {Aad}},
  \bibinfo {author} {\bibfnamefont {B.}~\bibnamefont {Abbott}}, \bibinfo
  {author} {\bibfnamefont {O.}~\bibnamefont {Abdinov}}, \bibinfo {author}
  {\bibfnamefont {B.}~\bibnamefont {Abeloos}}, \bibinfo {author} {\bibfnamefont
  {D.K.}\ \bibnamefont {Abhayasinghe}}, \bibinfo {author} {\bibfnamefont
  {S.H.}\ \bibnamefont {Abidi}}, \bibinfo {author} {\bibfnamefont {O.S.}\
  \bibnamefont {AbouZeid}}, \bibinfo {author} {\bibfnamefont {N.L.}\
  \bibnamefont {Abraham}}, \ and\ \bibinfo {author} {\bibnamefont {et~al.}},\
  }\bibfield  {title} {\enquote {\bibinfo {title} {{Measurement of $W^\pm Z$
  production cross sections and gauge boson polarisation in pp collisions at
  $\sqrt{s} = 13$ TeV with the ATLAS detector}},}\ }\href {\doibase
  10.1140/epjc/s10052-019-7027-6} {\bibfield  {journal} {\bibinfo  {journal}
  {The European Physical Journal C}\ }\textbf {\bibinfo {volume} {79}}
  (\bibinfo {year} {2019}),\ 10.1140/epjc/s10052-019-7027-6}\BibitemShut
  {NoStop}%
\bibitem [{\citenamefont {Panico}\ \emph {et~al.}(2018)\citenamefont {Panico},
  \citenamefont {Riva},\ and\ \citenamefont {Wulzer}}]{Riva:2018}%
  \BibitemOpen
  \bibfield  {author} {\bibinfo {author} {\bibfnamefont {G.}~\bibnamefont
  {Panico}}, \bibinfo {author} {\bibfnamefont {F.}~\bibnamefont {Riva}}, \ and\
  \bibinfo {author} {\bibfnamefont {A.}~\bibnamefont {Wulzer}},\ }\bibfield
  {title} {\enquote {\bibinfo {title} {{Diboson interference resurrection}},}\
  }\href {\doibase 10.1016/j.physletb.2017.11.068} {\bibfield  {journal}
  {\bibinfo  {journal} {Physics Letters B}\ }\textbf {\bibinfo {volume}
  {776}},\ \bibinfo {pages} {473–480} (\bibinfo {year} {2018})}\BibitemShut
  {NoStop}%
\bibitem [{\citenamefont {Azatov}\ \emph {et~al.}(2019)\citenamefont {Azatov},
  \citenamefont {Barducci},\ and\ \citenamefont {Venturini}}]{Barducci:2019}%
  \BibitemOpen
  \bibfield  {author} {\bibinfo {author} {\bibfnamefont {A.}~\bibnamefont
  {Azatov}}, \bibinfo {author} {\bibfnamefont {D.}~\bibnamefont {Barducci}}, \
  and\ \bibinfo {author} {\bibfnamefont {E.}~\bibnamefont {Venturini}},\
  }\bibfield  {title} {\enquote {\bibinfo {title} {{Precision diboson
  measurements at hadron colliders}},}\ }\href {\doibase
  10.1007/jhep04(2019)075} {\bibfield  {journal} {\bibinfo  {journal} {Journal
  of High Energy Physics}\ }\textbf {\bibinfo {volume} {2019}} (\bibinfo {year}
  {2019}),\ 10.1007/jhep04(2019)075}\BibitemShut {NoStop}%
\bibitem [{\citenamefont {Rahaman}\ and\ \citenamefont
  {Singh}(2020)}]{Rahaman:2020}%
  \BibitemOpen
  \bibfield  {author} {\bibinfo {author} {\bibfnamefont {R.}~\bibnamefont
  {Rahaman}}\ and\ \bibinfo {author} {\bibfnamefont {R.K.}\ \bibnamefont
  {Singh}},\ }\bibfield  {title} {\enquote {\bibinfo {title} {{Unravelling the
  anomalous gauge boson couplings in $Z W^\pm$ production at the LHC and the
  role of spin-1 polarizations}},}\ }\href {\doibase 10.48550/arXiv.1911.03111}
  {\bibfield  {journal} {\bibinfo  {journal} {Journal of High Energy Physics}\
  }\textbf {\bibinfo {volume} {2020}} (\bibinfo {year} {2020}),\
  10.48550/arXiv.1911.03111}\BibitemShut {NoStop}%
\bibitem [{\citenamefont {Grazzini}\ \emph
  {et~al.}(2017{\natexlab{a}})\citenamefont {Grazzini}, \citenamefont
  {Kallweit},\ and\ \citenamefont {Wiesemann}}]{Grazzini:2017mhc}%
  \BibitemOpen
  \bibfield  {author} {\bibinfo {author} {\bibfnamefont {M.}~\bibnamefont
  {Grazzini}}, \bibinfo {author} {\bibfnamefont {S.}~\bibnamefont {Kallweit}},
  \ and\ \bibinfo {author} {\bibfnamefont {M.}~\bibnamefont {Wiesemann}},\
  }\bibfield  {title} {\enquote {\bibinfo {title} {{Fully differential NNLO
  computations with MATRIX}},}\ }\href@noop {} {\  (\bibinfo {year}
  {2017}{\natexlab{a}})},\ \Eprint {http://arxiv.org/abs/1711.06631}
  {arXiv:1711.06631 [hep-ph]} \BibitemShut {NoStop}%
\bibitem [{\citenamefont {Gehrmann}\ \emph {et~al.}(2015)\citenamefont
  {Gehrmann}, \citenamefont {von Manteuffel},\ and\ \citenamefont
  {Tancredi}}]{Gehrmann:2015ora}%
  \BibitemOpen
  \bibfield  {author} {\bibinfo {author} {\bibfnamefont {T.}~\bibnamefont
  {Gehrmann}}, \bibinfo {author} {\bibfnamefont {A.}~\bibnamefont {von
  Manteuffel}}, \ and\ \bibinfo {author} {\bibfnamefont {L.}~\bibnamefont
  {Tancredi}},\ }\bibfield  {title} {\enquote {\bibinfo {title} {{The two-loop
  helicity amplitudes for $q\overline{q}^{\prime} \rightarrow V_1 V_2
  \rightarrow 4 $ leptons}},}\ }\href {\doibase 10.1007/JHEP09(2015)128}
  {\bibfield  {journal} {\bibinfo  {journal} {JHEP}\ }\textbf {\bibinfo
  {volume} {09}},\ \bibinfo {pages} {128} (\bibinfo {year} {2015})},\ \Eprint
  {http://arxiv.org/abs/1503.04812} {arXiv:1503.04812 [hep-ph]} \BibitemShut
  {NoStop}%
\bibitem [{\citenamefont {Denner}\ \emph {et~al.}(2017)\citenamefont {Denner},
  \citenamefont {Dittmaier},\ and\ \citenamefont {Hofer}}]{Denner:2016kdg}%
  \BibitemOpen
  \bibfield  {author} {\bibinfo {author} {\bibfnamefont {A.}~\bibnamefont
  {Denner}}, \bibinfo {author} {\bibfnamefont {S.}~\bibnamefont {Dittmaier}}, \
  and\ \bibinfo {author} {\bibfnamefont {L.}~\bibnamefont {Hofer}},\ }\bibfield
   {title} {\enquote {\bibinfo {title} {{Collier: a fortran-based Complex
  One-Loop LIbrary in Extended Regularizations}},}\ }\href {\doibase
  10.1016/j.cpc.2016.10.013} {\bibfield  {journal} {\bibinfo  {journal}
  {Comput. Phys. Commun.}\ }\textbf {\bibinfo {volume} {212}},\ \bibinfo
  {pages} {220--238} (\bibinfo {year} {2017})},\ \Eprint
  {http://arxiv.org/abs/1604.06792} {arXiv:1604.06792 [hep-ph]} \BibitemShut
  {NoStop}%
\bibitem [{\citenamefont {Cascioli}\ \emph {et~al.}(2012)\citenamefont
  {Cascioli}, \citenamefont {Maierhofer},\ and\ \citenamefont
  {Pozzorini}}]{Cascioli:2011va}%
  \BibitemOpen
  \bibfield  {author} {\bibinfo {author} {\bibfnamefont {F.}~\bibnamefont
  {Cascioli}}, \bibinfo {author} {\bibfnamefont {P.}~\bibnamefont
  {Maierhofer}}, \ and\ \bibinfo {author} {\bibfnamefont {S.}~\bibnamefont
  {Pozzorini}},\ }\bibfield  {title} {\enquote {\bibinfo {title} {{Scattering
  Amplitudes with Open Loops}},}\ }\href {\doibase
  10.1103/PhysRevLett.108.111601} {\bibfield  {journal} {\bibinfo  {journal}
  {Phys. Rev. Lett.}\ }\textbf {\bibinfo {volume} {108}},\ \bibinfo {pages}
  {111601} (\bibinfo {year} {2012})},\ \Eprint {http://arxiv.org/abs/1111.5206}
  {arXiv:1111.5206 [hep-ph]} \BibitemShut {NoStop}%
\bibitem [{\citenamefont {Buccioni}\ \emph {et~al.}(2019)\citenamefont
  {Buccioni}, \citenamefont {Lang}, \citenamefont {Lindert}, \citenamefont
  {Maierhofer}, \citenamefont {Pozzorini}, \citenamefont {Zhang},\ and\
  \citenamefont {Zoller}}]{Buccioni:2019sur}%
  \BibitemOpen
  \bibfield  {author} {\bibinfo {author} {\bibfnamefont {F.}~\bibnamefont
  {Buccioni}}, \bibinfo {author} {\bibfnamefont {J.-N.}\ \bibnamefont {Lang}},
  \bibinfo {author} {\bibfnamefont {J.M.}\ \bibnamefont {Lindert}}, \bibinfo
  {author} {\bibfnamefont {P.}~\bibnamefont {Maierhofer}}, \bibinfo {author}
  {\bibfnamefont {S.}~\bibnamefont {Pozzorini}}, \bibinfo {author}
  {\bibfnamefont {H.}~\bibnamefont {Zhang}}, \ and\ \bibinfo {author}
  {\bibfnamefont {M.F.}\ \bibnamefont {Zoller}},\ }\bibfield  {title} {\enquote
  {\bibinfo {title} {{OpenLoops 2}},}\ }\href {\doibase
  10.1140/epjc/s10052-019-7306-2} {\bibfield  {journal} {\bibinfo  {journal}
  {Eur. Phys. J. C}\ }\textbf {\bibinfo {volume} {79}},\ \bibinfo {pages} {866}
  (\bibinfo {year} {2019})},\ \Eprint {http://arxiv.org/abs/1907.13071}
  {arXiv:1907.13071 [hep-ph]} \BibitemShut {NoStop}%
\bibitem [{\citenamefont {Buccioni}\ \emph {et~al.}(2018)\citenamefont
  {Buccioni}, \citenamefont {Pozzorini},\ and\ \citenamefont
  {Zoller}}]{Buccioni:2017yxi}%
  \BibitemOpen
  \bibfield  {author} {\bibinfo {author} {\bibfnamefont {F.}~\bibnamefont
  {Buccioni}}, \bibinfo {author} {\bibfnamefont {S.}~\bibnamefont {Pozzorini}},
  \ and\ \bibinfo {author} {\bibfnamefont {M.}~\bibnamefont {Zoller}},\
  }\bibfield  {title} {\enquote {\bibinfo {title} {{On-the-fly reduction of
  open loops}},}\ }\href {\doibase 10.1140/epjc/s10052-018-5562-1} {\bibfield
  {journal} {\bibinfo  {journal} {Eur. Phys. J. C}\ }\textbf {\bibinfo {volume}
  {78}},\ \bibinfo {pages} {70} (\bibinfo {year} {2018})},\ \Eprint
  {http://arxiv.org/abs/1710.11452} {arXiv:1710.11452 [hep-ph]} \BibitemShut
  {NoStop}%
\bibitem [{\citenamefont {Catani}\ \emph {et~al.}(2012)\citenamefont {Catani},
  \citenamefont {Cieri}, \citenamefont {de~Florian}, \citenamefont {Ferrera},\
  and\ \citenamefont {Grazzini}}]{Catani:2012qa}%
  \BibitemOpen
  \bibfield  {author} {\bibinfo {author} {\bibfnamefont {S.}~\bibnamefont
  {Catani}}, \bibinfo {author} {\bibfnamefont {L.}~\bibnamefont {Cieri}},
  \bibinfo {author} {\bibfnamefont {D.}~\bibnamefont {de~Florian}}, \bibinfo
  {author} {\bibfnamefont {G.}~\bibnamefont {Ferrera}}, \ and\ \bibinfo
  {author} {\bibfnamefont {M.}~\bibnamefont {Grazzini}},\ }\bibfield  {title}
  {\enquote {\bibinfo {title} {{Vector boson production at hadron colliders:
  hard-collinear coefficients at the NNLO}},}\ }\href {\doibase
  10.1140/epjc/s10052-012-2195-7} {\bibfield  {journal} {\bibinfo  {journal}
  {Eur. Phys. J.}\ }\textbf {\bibinfo {volume} {C72}},\ \bibinfo {pages} {2195}
  (\bibinfo {year} {2012})},\ \Eprint {http://arxiv.org/abs/1209.0158}
  {arXiv:1209.0158 [hep-ph]} \BibitemShut {NoStop}%
\bibitem [{\citenamefont {Catani}\ and\ \citenamefont
  {Grazzini}(2007)}]{Catani:2007vq}%
  \BibitemOpen
  \bibfield  {author} {\bibinfo {author} {\bibfnamefont {S.}~\bibnamefont
  {Catani}}\ and\ \bibinfo {author} {\bibfnamefont {M.}~\bibnamefont
  {Grazzini}},\ }\bibfield  {title} {\enquote {\bibinfo {title} {{An NNLO
  subtraction formalism in hadron collisions and its application to Higgs boson
  production at the LHC}},}\ }\href {\doibase 10.1103/PhysRevLett.98.222002}
  {\bibfield  {journal} {\bibinfo  {journal} {Phys. Rev. Lett.}\ }\textbf
  {\bibinfo {volume} {98}},\ \bibinfo {pages} {222002} (\bibinfo {year}
  {2007})},\ \Eprint {http://arxiv.org/abs/hep-ph/0703012}
  {arXiv:hep-ph/0703012 [hep-ph]} \BibitemShut {NoStop}%
\bibitem [{\citenamefont {Grazzini}\ \emph
  {et~al.}(2017{\natexlab{b}})\citenamefont {Grazzini}, \citenamefont
  {Kallweit}, \citenamefont {Rathlev},\ and\ \citenamefont
  {Wiesemann}}]{Grazzini:2017ckn}%
  \BibitemOpen
  \bibfield  {author} {\bibinfo {author} {\bibfnamefont {M.}~\bibnamefont
  {Grazzini}}, \bibinfo {author} {\bibfnamefont {S.}~\bibnamefont {Kallweit}},
  \bibinfo {author} {\bibfnamefont {D.}~\bibnamefont {Rathlev}}, \ and\
  \bibinfo {author} {\bibfnamefont {M.}~\bibnamefont {Wiesemann}},\ }\bibfield
  {title} {\enquote {\bibinfo {title} {{$W^\pm Z$ production at the LHC:
  fiducial cross sections and distributions in NNLO QCD}},}\ }\href@noop {} {\
  (\bibinfo {year} {2017}{\natexlab{b}})},\ \Eprint
  {http://arxiv.org/abs/1703.09065} {arXiv:1703.09065 [hep-ph]} \BibitemShut
  {NoStop}%
\bibitem [{\citenamefont {Grazzini}\ \emph {et~al.}(2016)\citenamefont
  {Grazzini}, \citenamefont {Kallweit}, \citenamefont {Rathlev},\ and\
  \citenamefont {Wiesemann}}]{Grazzini:2016swo}%
  \BibitemOpen
  \bibfield  {author} {\bibinfo {author} {\bibfnamefont {M.}~\bibnamefont
  {Grazzini}}, \bibinfo {author} {\bibfnamefont {S.}~\bibnamefont {Kallweit}},
  \bibinfo {author} {\bibfnamefont {D.}~\bibnamefont {Rathlev}}, \ and\
  \bibinfo {author} {\bibfnamefont {M.}~\bibnamefont {Wiesemann}},\ }\bibfield
  {title} {\enquote {\bibinfo {title} {{$W^{\pm}Z$ production at hadron
  colliders in NNLO QCD}},}\ }\href {\doibase 10.1016/j.physletb.2016.08.017}
  {\bibfield  {journal} {\bibinfo  {journal} {Phys. Lett.}\ }\textbf {\bibinfo
  {volume} {B761}},\ \bibinfo {pages} {179--183} (\bibinfo {year} {2016})},\
  \Eprint {http://arxiv.org/abs/1604.08576} {arXiv:1604.08576 [hep-ph]}
  \BibitemShut {NoStop}%
\bibitem [{\citenamefont {Azatov}\ \emph
  {et~al.}(2017{\natexlab{b}})\citenamefont {Azatov}, \citenamefont
  {Elias-Miró}, \citenamefont {Reyimuaji},\ and\ \citenamefont
  {Venturini}}]{NovelTGC:2017}%
  \BibitemOpen
  \bibfield  {author} {\bibinfo {author} {\bibfnamefont {A.}~\bibnamefont
  {Azatov}}, \bibinfo {author} {\bibfnamefont {J.}~\bibnamefont {Elias-Miró}},
  \bibinfo {author} {\bibfnamefont {Y.}~\bibnamefont {Reyimuaji}}, \ and\
  \bibinfo {author} {\bibfnamefont {E.}~\bibnamefont {Venturini}},\ }\bibfield
  {title} {\enquote {\bibinfo {title} {{Novel measurements of anomalous triple
  gauge couplings for the LHC}},}\ }\href {\doibase 10.1007/jhep10(2017)027}
  {\bibfield  {journal} {\bibinfo  {journal} {Journal of High Energy Physics}\
  }\textbf {\bibinfo {volume} {2017}} (\bibinfo {year} {2017}{\natexlab{b}}),\
  10.1007/jhep10(2017)027}\BibitemShut {NoStop}%
\bibitem [{\citenamefont {Baglio}\ \emph {et~al.}(2019)\citenamefont {Baglio},
  \citenamefont {Dawson},\ and\ \citenamefont {Homiller}}]{Dawson:2019}%
  \BibitemOpen
  \bibfield  {author} {\bibinfo {author} {\bibfnamefont {J.}~\bibnamefont
  {Baglio}}, \bibinfo {author} {\bibfnamefont {S.}~\bibnamefont {Dawson}}, \
  and\ \bibinfo {author} {\bibfnamefont {S.}~\bibnamefont {Homiller}},\
  }\bibfield  {title} {\enquote {\bibinfo {title} {{QCD corrections in Standard
  Model EFT fits to $WZ$ and $WW$ production}},}\ }\href {\doibase
  10.1103/physrevd.100.113010} {\bibfield  {journal} {\bibinfo  {journal}
  {Physical Review D}\ }\textbf {\bibinfo {volume} {100}} (\bibinfo {year}
  {2019}),\ 10.1103/physrevd.100.113010}\BibitemShut {NoStop}%
\bibitem [{\citenamefont {Bern}\ \emph {et~al.}(2011)\citenamefont {Bern},
  \citenamefont {Diana}, \citenamefont {Dixon}, \citenamefont {Cordero},
  \citenamefont {Forde}, \citenamefont {Gleisberg}, \citenamefont {Höche},
  \citenamefont {Ita}, \citenamefont {Kosower}, \citenamefont {Maître},\ and\
  \citenamefont {et~al.}}]{Dixon:2011}%
  \BibitemOpen
  \bibfield  {author} {\bibinfo {author} {\bibfnamefont {Z.}~\bibnamefont
  {Bern}}, \bibinfo {author} {\bibfnamefont {G.}~\bibnamefont {Diana}},
  \bibinfo {author} {\bibfnamefont {L.J.}\ \bibnamefont {Dixon}}, \bibinfo
  {author} {\bibfnamefont {F.F.}\ \bibnamefont {Cordero}}, \bibinfo {author}
  {\bibfnamefont {D.}~\bibnamefont {Forde}}, \bibinfo {author} {\bibfnamefont
  {T.}~\bibnamefont {Gleisberg}}, \bibinfo {author} {\bibfnamefont
  {S.}~\bibnamefont {Höche}}, \bibinfo {author} {\bibfnamefont
  {H.}~\bibnamefont {Ita}}, \bibinfo {author} {\bibfnamefont {D.A.}\
  \bibnamefont {Kosower}}, \bibinfo {author} {\bibfnamefont {D.}~\bibnamefont
  {Maître}}, \ and\ \bibinfo {author} {\bibnamefont {et~al.}},\ }\bibfield
  {title} {\enquote {\bibinfo {title} {{Left-handed $W$ bosons at the LHC}},}\
  }\href {\doibase 10.1103/physrevd.84.034008} {\bibfield  {journal} {\bibinfo
  {journal} {Physical Review D}\ }\textbf {\bibinfo {volume} {84}} (\bibinfo
  {year} {2011}),\ 10.1103/physrevd.84.034008}\BibitemShut {NoStop}%
\bibitem [{\citenamefont {Collaboration}(2022)}]{WgCMS:2021}%
  \BibitemOpen
  \bibfield  {author} {\bibinfo {author} {\bibfnamefont {CMS}\ \bibnamefont
  {Collaboration}},\ }\bibfield  {title} {\enquote {\bibinfo {title} {{$W^\pm
  \gamma$ differential cross sections and effective field theory constraints at
  $\sqrt{s}=13$ TeV}},}\ }\href {\doibase
  https://doi.org/10.1103/PhysRevD.105.052003} {\bibfield  {journal} {\bibinfo
  {journal} {Phys. Rev. D}\ }\textbf {\bibinfo {volume} {105}} (\bibinfo {year}
  {2022}),\ https://doi.org/10.1103/PhysRevD.105.052003}\BibitemShut {NoStop}%
\end{thebibliography}%
\end{document}